\documentclass[hidelinks,onefignum,onetabnum]{siamart251216}

\usepackage{amssymb}
\usepackage{amsmath}
\usepackage{booktabs}
\usepackage{mathtools}
\usepackage{enumerate}
\usepackage{float}
\usepackage{graphicx} 
\graphicspath{ {./figures_new/} }    
\usepackage{tabularx}
\usepackage{tikz}
\usetikzlibrary{quantikz2}  
\usepackage{amsopn}
\DeclareMathOperator{\diag}{diag}

\usepackage{xr-hyper}
\usepackage{hyperref}               
\hypersetup{
	colorlinks=true,
	linkcolor=red,
	filecolor=magenta,
	urlcolor=blue
	}

\usepackage{cleveref}

\ifpdf
  \DeclareGraphicsExtensions{.eps,.pdf,.png,.jpg}
\else
  \DeclareGraphicsExtensions{.eps}
\fi

\newsiamremark{remark}{Remark}
\newsiamremark{hypothesis}{Hypothesis}
\crefname{hypothesis}{Hypothesis}{Hypotheses}
\newsiamthm{example}{Example}
\newsiamthm{problem}{Problem}
\newsiamremark{fact}{Fact}
\crefname{fact}{Fact}{Facts}

\crefname{equation}{eq.}{eqs.}
\Crefname{equation}{Eq.}{Eqs.}

\headers{An Example Article}{D. Doe, P. T. Frank, and J. E. Smith}

\title{A quantum advection-diffusion solver using the quantum singular value transform}

\author{Gard Olav Helle\thanks{Center for Quantum Mathematics, University of Southern Denmark, 5230 Odense, Denmark 
  (\email{gardoh@imada.sdu.dk})}
\and Tommaso Benacchio\thanks{Weather Research, Danish Meteorological Institute, 2100 Copenhagen, Denmark}
\and Anna Bomme Ousager\footnotemark[1]
\and Jørgen Ellegaard Andersen\footnotemark[1]}

\newcommand{\C}{\mathbb{C}}
\newcommand{\R}{\mathbb{R}}
\newcommand{\Z}{\mathbb{Z}}
\newcommand{\N}{\mathbb{N}}

\newcommand{\MC}[1]{\mathcal{#1}}
\newcommand{\ra}{\rightarrow}
\newcommand{\om}[1]{\operatorname{#1}}
\newcommand{\dd}{\partial}
\newcommand{\eps}{\epsilon}

\headers{}{}

\begin{document}

\maketitle

\begin{abstract}
 We present a quantum algorithm for the simulation of the linear advection-diffusion equation based on block encodings of high order finite-difference operators and the quantum singular value transform. Our complexity analysis shows that the higher order methods significantly reduce the number of gates and qubits required to reach a given accuracy. The theoretical results are supported by numerical simulations of one- and two-dimensional benchmarks.   
\end{abstract}

\begin{keywords}
high-order finite-difference method, advection-diffusion equation, quantum algorithm, quantum singular value transform, block-encoding, complexity analysis
\end{keywords}

\begin{MSCcodes}
68Q12, 65M15, 65M22
\end{MSCcodes}

\section{Introduction}
Computational fluid dynamics simulates and analyses fluid flow using numerical approximations. This involves solving non-linear partial differential equation, typically some incarnation of the Navier-Stokes equation, and requires vast computational resources. Advanced applications include aerospace and automotive engineering and weather prediction. In the latter, while state-of-the-art numerical models have made remarkable progress in the last decades, the feasibility of further upgrades clashes with the reality of energy consumption of high-performance computing facilities that are ever increasing in size to match the challenge of simulations at higher and higher resolutions \cite{Bauer_etal15}. Significant efficiency improvements are necessary, questioning the sustainability of incremental upgrades of legacy methods \cite{Schulthess_etal18}.  

Quantum computers may provide a new avenue for supplying next-generation efficiency in improved fluid dynamical simulations. Numerous quantum algorithms have been developed for the solution of ordinary and partial differential equations (ODEs and PDEs) \cite{Berry_etal17, Krovi23, ChildsOstrander21,AnChildsLin23,BerryCosta22}. The archetypal example is the Schr\"{o}dinger equation, which is known to be efficiently solvable on a quantum computer for several families of Hamiltonians \cite{LC17}. It has been demonstrated that non-unitary dynamics can be effectively simulated in certain cases \cite{Liu_etal21}, but obstacles towards generalization remain. A well-known limitation is the presence of exponential growth or decay in the simulated dynamics. In a quantum computer, this situation must be handled via rescaling and post-selection, which is infeasible for complexity theoretical reasons \cite{Aaronson04_PostBQP}.

The prevailing approach to simulating PDEs and ODEs is to discretize the problem into a linear system, using finite-difference, finite element, finite volume or spectral methods. The system is then solved using a state-of-the-art quantum linear systems algorithm \cite{HHL09, Childs_Kothari_Somma15, QSVT19}. Many time-dependent PDEs can be transformed into a system of ODEs by discretizing the problem in space first. The resulting time-dependent problem can then be solved using specialized quantum algorithms. This is the strategy adopted in this work. Due to the linear nature of quantum computers, non-linear equations must be handled through some notion of linearization scheme. Popular approaches include Carleman linearization \cite{Liu_etal21,Krovi23,Costa_etal25, Jennings_etal25a}, Koopman von Neumann quantization \cite{Joseph20} and the homotopy analysis method \cite{Xue_etal25}.

In this paper, we present a quantum algorithm that implements a numerical solution, based on high-order finite-difference approximations, to the advection-diffusion equation 
\begin{equation} \label{eq:adv_diff_general}
\dd_t u + c\cdot \nabla u  = \nu \Delta u \quad \text{ for } \quad u\colon [0,T]\times [0,d]^n \ra \R, 
\end{equation} 
with periodic boundary conditions, where $c\in \R^n$ is the advection speed, $\nu$ is the molecular diffusivity, $\nabla$ is the gradient, and $\Delta$ the Laplacian. We focus on the one- and two-dimensional cases and present the construction in terms of one- and two-qubit gates in detail. This results in an easily implementable algorithm with finite-difference approximations of order $2$, $4$, $6$ and $14$ that can be tested on current simulators and hardware.  

\subsection{Summary of main results}
To outline our numerical approach, consider a general partial differential equation of the form 
\begin{equation} 
\dd_t u(t,x) = Pu(t,x)  \quad u\colon [0,T]\times [0,d]^n \ra \R, 
\end{equation}
where $P$ a constant coefficient spatial differential operator. Let $L$ be a finite-difference approximation to $P$ and consider the associated difference-differential equation
\begin{equation} \label{Difference-Differential-eq}
    \dot{v}(t) = Lv(t) \quad \text{ for } v\colon [0,T]\ra \R^N, 
\end{equation}
where $v(t)\in \R^N$ represents an approximation of $u(t,x)$ on $N$ uniformly distributed grid points. Our approach is to prepare
an approximation of the solution $v(t) = e^{Lt}v_0$ of \cref{Difference-Differential-eq} using the quantum singular value transform (QSVT) \cite{QSVT19}. There are two basic inputs to the QSVT algorithm: a block-encoding and a polynomial. For the first input, we construct a block-encoding of $H = i\beta D_{2p}$, where $D_{2p} \approx \dd_x$ is a symmetric finite-difference operator of order $2p\geq 2$ and $\beta$ an appropriate normalizing constant. For the second input, we construct a polynomial approximation of $e^{-M_1x^2 + iM_2x}$ of any given accuracy $\eps>0$. The basic idea of the block-encoding is that any finite-difference operator is a linear combination of translation operators, so an incarnation of the linear combination of unitaries (LCU) method \cite{Childs_Wiebe12} can be utilized. A generalization to block-encodings of pseudo-differential operators is given in \cite{LiNiYing23}. 

Our main contribution is the construction and careful analysis of a quantum advection-diffusion algorithm in the block-encoding plus QSVT framework, with a particular emphasis on the scaling of high-order methods. On the one hand, we rigorously express the gate complexity and qubit requirements in terms of basic problem parameters. On the other hand, our numerical demonstrations show that the algorithm can be effectively implemented with a comparatively small set of gates while achieving high precision results. Our complexity analysis includes a precise error estimate for the solution of \cref{Difference-Differential-eq} given in \cref{thm:L2_estimate} and a degree bound on the polynomial approximation needed to achieve a given precision using QSVT \cref{prop:product_approx}. Along the way
we also establish precise error estimates for the eigenvalues of $D_{2p}$ and the second order analogue $D_{2p}^{(2)}$ (used in both \cite{ChildsOstrander21} and \cite{Kivlichan_2017}) in \cref{thm:lambda_mu_estimates}.

The problem we consider is formulated precisely in \cref{prob:quantum_adv_diff}:  Given an evolution time $T>0$, a precision $\eps>0$ and an amplitude encoding of the initial function $u_0(x)$, prepare an amplitude encoding of the exact solution $u_T$ of \cref{eq:adv_diff_general} with precision $\eps>0$. Our complexity results are given in \cref{thm:main_complexity,cor:complexity_simplified}. In the presence of some mild assumptions laid out in \cref{sec:complexity}, we have the following special cases. For pure advection, $\nu = 0$, the algorithm uses $n+m+2$ qubits, where 
\[ n = \lceil \log_2(d(cT||u_0^{(2p+1)}||_{L^2}/(\eps||u_0||_{L^2}))^{1/(2p)}) \rceil \qquad \text{ and } \qquad m = \lceil \log_2(2p+1) \rceil \]  
and the gate complexity is given by
\begin{equation} \label{eq:complexity_advection} 
\widetilde{O}\left((cT)^{1+\frac1{2p}}\left(\frac{||u_0^{(2p+1)}||_{L^2}}{||u_0||_{L^2}}\right)^{\frac{1}{2p}}\eps^{-\frac{1}{2p}}\log(p)^2\right)        
\end{equation}
For pure diffusion, $c = 0$, the qubit requirements are similar and the gate complexity is given by 
\begin{equation} \label{eq:complexity_diffusion} 
\widetilde{O}\left(\sqrt{(\nu T)^{1+\frac1{p}} e^{-\nu T \mu_1'/p}}\left(\frac{||u_0^{(2p+2)}||_{L^2}}{||u_0||_{L^2}}\right)^{\frac{1}{2p}}\eps^{-\frac{1}{2p}}\log(p)^2\right).      
\end{equation}
These formulas show that the dependence on precision $\eps$ improves dramatically with the order $2p$ of the finite-difference approximations, while the time dependence becomes closer to linear in $cT$ for pure advection and square-root depends on $\nu T$ for pure diffusion. The main cost of increasing $p$ is the final logarithmic factor.

\subsection{Related work}
Several quantum algorithms have been proposed for the simulation of the advection-diffusion dynamics. 
In \cite{Ingelmann_etal24} the PDE was reduced to a linear system using low order finite-difference approximations and tackled using the HHL algorithm. An approach based on low order finite differences and time-marching was studied in \cite{BrearleySylvain24,Over_etal25}. More exotic methods linear combination of Hamiltonian simulation \cite{Novikau24}, Schr\"{o}dingerization \cite{Hu_etal24} and probabilistic imaginary time evolution \cite{Xie_etal24}. In \cite{Lubasch_etal25} the authors developed explicit circuits for the advection, heat, wave, and Poisson equations based on ideas similar to those of the current work, that is, block encoding and quantum singular value transform in Fourier space. However, to our knowledge, a careful study of the implementation and performance of high-order finite-difference operators in this context has not yet been carried out.

\subsection{Organization}
The paper is structured as follows. In \cref{sec:background} we provide background on the advection-diffusion equation, finite-difference operators and QSVT. In \cref{sec:block-encoding}, we construct the block-encoding of $D_{2p}$. \Cref{sec:complexity} is devoted to analyzing the complexity of the algorithm involving both numerical estimates and the construction of appropriate polynomial approximations. In \cref{sec:high-dim_case}, we explain how the $1d$ algorithm can be used to handle the general high-dimensional case. In \cref{sec:numerical_results}, we display and discuss numerical results. The final section draws conclusions and gives an outlook for future work.

\section{Background and problem formulation}  \label{sec:background}
We restrict our attention to the one-dimensional advection-diffusion equation 
\begin{equation}  \label{eq:adv_diff}
\dd_tu+c\dd_xu  = \nu \dd_x^2u \quad \text{ for } \quad u\colon [0,T] \times [0,d]\ra \R 
\end{equation} 
with periodic boundary conditions. For $k\in \Z$ define
\begin{equation} \label{eq:ek_omega_def}
e_k(x)\coloneqq e^{i\omega kx} \quad \text{ where } \omega = 2\pi/d.
\end{equation}
If the initial function is expanded in a Fourier series, $u_0(x) = \sum_{k\in \Z} a_k e_k(x)$,
then the exact solution of \cref{eq:adv_diff} is given by
\begin{equation} \label{eq:exact_solution}
u(t,x) = \sum_{k\in \Z} a_ke^{-\nu \omega^2k^2t}e^{-ic\omega kt}e_k(x).
\end{equation} 
Note that $u_0$ being of class $C^1$ is sufficient to ensure absolute convergence of the Fourier series. In the case of pure advection, $\nu = 0$, the exact solution takes the simpler form $u(t,x) = u_0(x-ct)$. One could attempt to prepare the exact solution directly, but this is not the goal of our study. Instead, the aim is to develop techniques for efficiently encoding numerical methods with a view towards more complex problems.

\begin{problem} \label{prob:quantum_adv_diff}
Let $u_0\colon [0,d]\ra \R$ be a smooth, periodic initial function, $T>0$ the evolution time, $\eps>0$ the desired precision and $N_0>0$ the minimal number of grid points. Let $u(t,x)$ be the exact solution of \cref{eq:adv_diff} and for a given $N = 2^n\geq N_0$ let 
\begin{equation} \label{eq:initial_state}
\ket{\bar{u}_t} = \beta_0^{-1}\sum_{j=0}^{N-1} u_0(t,j\Delta x)\ket{j},   
\end{equation} 
where $\beta_0$ is a normalizing constant ensuring that $||\ket{\bar{u}_0}|| = 1$. Assuming access to $\ket{\bar{u}_0}$, prepare a quantum state $\ket{\bar{w}_T}$ such that 
$||\ket{\bar{u}_T}-\ket{\bar{w}_T} ||\leq \eps $.
\end{problem}

\begin{remark} Note that we do not require $\ket{\bar{u}_T}$ and $\ket{\bar{w}_T}$ to have unit norm. The corresponding normalized quantum state can be obtained via amplitude amplitude amplification using $O(1/||\bar{u}_T||)$ queries to the state preparation unitary and our quantum algorithm. 
\end{remark} 

The number $N_0$ should be understood as the minimal number of grid points or Fourier modes needed to faithfully represent the initial function $u_0$. In fact, in our analysis we will assume that $u_0(x)= \sum_{k=-(N_0-1)/2}^{(N_0-1)/2} a_k e_k(x)$, which is reasonable since errors made in the initial representation propagate at worst linearly (see \cref{lem:error_propagation}). 

\subsection{Background on QSVT} 
Throughout the paper, let $H_n \coloneqq \C^{2^n}$, $n\geq 1$, denote the $n$-qubit Hilbert space and
$\{\ket{k}\colon 0\leq k<2^n\}$ the computational basis. The isomorphisms $(\C^2)^{\otimes n}\cong \C^{2^n}$ are specified by the least significant bit first convention, that is, 
$\ket{j_0}\ket{j_1}\cdots \ket{j_{n-1}}  \mapsto \ket{j}$ where $j \coloneqq \sum_{s=0}^{n-1}j_s2^s$. We follow the standard conventions on quantum gates and circuits given in \cite[chap.~4]{NielsenChuang10} unless otherwise stated. 
In particular, the standard Pauli gates are denoted by $X$, $Y$, $Z$ and the Hadamard gate is denoted by $H$. 

 A unitary block-encoding of a linear map $A\colon H_k\ra H_k$ is a pair
\[ (\iota\colon H_k\ra H_n,U_A\colon H_n\ra H_n),   \]
where $\iota$ is an isometry and $U$ is a unitary such that $A = \iota^\dagger \circ U\circ \iota \colon H_k\ra H_k$. We often let $\iota $ be known from context and specify the internal projection $\Pi = \iota \iota^\dagger\colon H_n\ra H_n$ instead. A block-encoding of $A$ exists if and only if the operator norm $||A||\leq 1$. If this condition fails, one must work with a scaled version $A/\alpha$ and keep track of the scaling factor $\alpha$.

The quantum singular value transform (QSVT) is a primitive that enables the manipulation of the singular values of a block-encoded matrix. Let $A\colon H_k \ra H_k$ be a matrix with singular value decomposition $A = U\diag(\sigma_j)_j V^\dagger$ and let $q(x)$ be a real polynomial of definite parity. Then the singular value transformation of $A$ by $q$ is defined as
$q^{SV}(A) \coloneqq U\diag(q(\sigma_j))_jV^\dagger$ if $q$ is odd and $q^{SV}(A) \coloneqq V\diag(q(\sigma_j))_jV^\dagger$ if $q$ is even \cite[Def.~16]{QSVT19}. Note that for a Hermitian matrix $H$, the singular value transformation coincides with the ordinary eigenvalue transformation $q^{SV}(H) = q(H)$. 

\begin{theorem}[QSVT informal description] \label{thm:QSVT_informal}
\item[\textbf{Input 1:}] A unitary block-encoding $(U\colon H_{a+n}\ra H_{a+n},\Pi)$ of $A\colon H_n\ra H_n$
\item[\textbf{Input 2:}] An even or odd real polynomial $q(x)\in \R[x]$ of degree $d$ satisfying $|q(x)|\leq 1$ for $x\in [-1,1]$. 
\item[\textbf{Output:}] A unitary block-encoding of $q(A)\colon H_n\ra H_n$
\[ (\om{QSVT}(U,\Pi,q) \colon H_{1+a+n}\ra H_{1+a+n},\ket{0}\bra{0}\otimes \Pi). \]
\item[\textbf{Cost:}] $1+a+n$ qubits, $d= \om{deg}q$ queries to the block-encoding, $2d$ queries to $C_\Pi\om{NOT} = \ket{0}\bra{0}\otimes \Pi + \ket{1}\bra{1}\otimes I$ and $d+2$ one-qubit gates.
\end{theorem}
\begin{remark} In practice, the polynomial $q(x)$ must be converted to an angle sequence that goes into the circuit construction. Additional details on this and precise formulations of the versions of QSVT we are using are given in \cref{appendix:QSVT}. \end{remark} 

The most common choice of isometry is $\iota\colon \ket{0}\otimes I\colon H_k \ra H_a\otimes H_k \cong H_{a+k}$, in which case, $\Pi = \ket{0}\bra{0}\otimes I$ and the $C_\Pi\om{NOT}$ gate is simply a multi-controlled $\om{NOT}$-gate with control state $\ket{0}\in H_a$. As long as the gate complexity of the block-encoding is greater than $O(a^2)$, which is the cost of an ancilla free implementation of the multi-controlled $\om{NOT}$-gate with $a$ controls, the gate complexity will be dominated by the $d$ queries to the block-encoding. 

The parity constraint on the polynomial $p$ can be overcome by applying the algorithm with the even and odd parts in parallel and combining the results with the linear combination of block-encodings method \cite[Sec.~ 10.2]{QAlg25}. Additional details on this and the quantum circuits are contained in the supplementary material.

\section{Numerical method} \label{sec:numerical_method}
As outlined in the introduction, our strategy is to replace the differential operator $P = -c\dd_x + \nu \dd_x^2$ by a finite-difference approximation. Introduce $N$ uniformly distributed grid points $x_j = j\Delta x$ for $0\leq j< N$, where $\Delta x \coloneqq d/N$. We adopt the convention $x_{j+Nk} \coloneqq x_j$ for $0\leq j<N$ and $k\in \Z$. For $j\in \N$ let $\delta_j$ be the symmetric finite difference operator
\begin{equation}  \label{eq:basic_FD}
\delta_jf(x) = \frac{f(x+j\Delta x/2)-f(x-j\Delta x/2)}{j\Delta x}
\end{equation} 
Symmetric finite-difference operators for the first and second derivatives of arbitrary orders are given in the following theorem. It can be derived from the more general results of \cite[Thm.~2.1,2.2]{Li05_FD}.

\begin{theorem} \label{thm:FD_operator}
For $p\geq 1$ the finite difference operators
\[ D_{2p} \coloneqq \sum_{j=1}^p \alpha_j \delta_{2j} \quad \text{ and } \quad \sum_{j=1}^pD_{2p}^{(2)}\coloneqq \alpha_j \delta_j^2 \quad
\text{ where } \alpha_j = \frac{2(-1)^{j+1}(p!)^2}{(p+j)!(p-j)!} \]
are accurate of order $2p$. More precisely, there are constants $C$ and $C'$ such that for all $f$ of class $C^{2p+1}$ and $g$ of class $C^{2p+2}$ one has
\begin{align*}
| f'(x)-D_{2p}f(x)| &\leq C ||f^{(2p+1)}||_{\infty,[x-p\Delta x,x+p\Delta x]} (\Delta x)^{2p}  \\
| g''(x)-D^{(2p)}_{2p}f(x)| &\leq C' ||f^{(2p+2)}||_{\infty,[x-p\Delta x,x+p\Delta x]} (\Delta x)^{2p}
\end{align*}
for all $x$ for which $[x-p\Delta x,x+p\Delta x]$ is contained in the domain of $f$ and $g$, respectively. 
\end{theorem} 

For an integer $p\geq 1$, we will consider the following two finite difference approximations
\begin{equation} \label{eq:FD_approx}
L = -cD_{2p}+\nu D_{2p}^2  \quad \text{ and } \quad  L = -cD_{2p}+\nu D_{2p}^{(2)}
\end{equation} 
of $P = -c\dd_x+\nu \dd_{x}^2$, both of order $2p$. This leads to the linear ODE $\dot{v}(t) = Lv(t)$ subject to $v(0) = (u_0(x_j))_{0\leq j<N}$ with solution $v(t) = e^{Lt}v_0$.
For the first choice of $L$ in \cref{eq:FD_approx} our strategy is to block-encode the Hermitian matrix $H = i\beta D_{2p}$, for a normalizing constant $\beta$ ensuring that $||H|| \leq 1$, and apply QSVT with $q$ a polynomial approximation of the function
\begin{equation}    \label{eq:exp_QSVT}
f(x;M_1,M_2) = e^{-M_1x^2+iM_2x} = e^{-M_1x^2}\cos(M_2x)+ie^{-M_1x^2}\sin(M_1x) ,
\end{equation}   
where $M_1 = cT/\beta$ and $M_2 = \nu T/\beta^2$. Hence $q(H)\approx f(H) = e^{LT}$, where the quality of the approximation is controlled by 
$\sup_{x\in [-1,1]} |f(x;M_1,M_2)-q(x)|$. This will be our primary choice of $L$ in the rest of the paper. 

Consider now the second choice $L = -cD_{2p}+\nu D_{2p}^{(2)}$ in \cref{eq:FD_approx}. Let $S = \nu D_{2p}^{(2)}$ and $H = icD_{2p}$. Then $S$ and $H$ are commuting Hermitian operators, so in particular $e^{Lt} = e^{St}e^{iHt}$. Here, both factors can be implemented with QSVT provided we have block-encodings of normalized versions of $S$ and $H$ and polynomial approximations of $g(x;M) = e^{Mx}$ and $f(x;0,M)$. The second factor corresponds to the previous situation with $f(x;0,M)$, but the first factor requires a polynomial approximation of $e^{Mx}$ on $[-1,0]$ (since the spectrum of $S$ is negative, see \cref{eq:def_lambda_mu}) that remains bounded by $1$ in $[0,1]$. In our experience, this approximation problem is more expensive to handle in practice. In summary, the second choice of $L$ requires two applications of QSVT with distinct block-encodings. This introduces an additional factor $1/2$ in the success probability that must be accounted for and there is an additional cost of composing the two block-encodings 
(see \cite[sec.~10.2]{QAlg25}). 

\section{Block encoding of finite-difference operators} \label{sec:block-encoding}
The goal of this section is to construct a block-encoding of the finite-difference operator
$D_{2p}$ given in \cref{thm:FD_operator}. More precisely, we will block-encode $H = i\beta D_{2p}$ for a suitable normalizing constant to be specified. The methods can be easily adapted to construct a block-encoding of $D_{2p}^{(2)}\approx \dd_x^2$ or any other finite difference operator. 

The key observation is that $D_{2p}$ is a linear combination of the basic finite difference operators $\delta_{2j}$ (see \cref{eq:basic_FD}) which in turn are linear combinations of translation operators. Since we work in a periodic domain, the resulting matrices are circulant and sparse. Efficient block encoding strategies for circulant matrices have been known for some time \cite{ZhouWang17} (see also the more recent \cite{Camps_etal24} and \cite[Sec.~V]{Motlagh_Wiebe24}). The basic idea is to use a modification of a modular adder \cite{Draper00} and the linear combination of unitaries (LCU) method \cite{Childs_Wiebe12}. 

In the following, it will be convenient to extend the notation $\ket{k}\in H_n$ to all $k\in \Z$ under the convention that
$\ket{k} = \ket{l}$ if and only if $k\equiv l \mod{2^n}$.
\begin{definition} \label{def:T_operator}
Define $T\in U(H_n)$ to be the translation operator given by $T\ket{k} = \ket{k+1}$ for all $k\in \Z$.
\end{definition} 

Note that $T$ has order $N = 2^n$ so that $T^{-1} = T^{N-1}$. Now, as operators on $H_n$ we have
\begin{equation}
    \delta_{2j}= \frac{1}{2j\Delta x}(T^j-T^{-j})  \quad \text{ and } \quad 
    \delta_j^2= \frac{1}{(j\Delta x)^2}(T^j-2I+T^{-j}).
\end{equation}
More generally, any linear combination of these operators can be expressed as a polynomial in $T$. The circulant matrices are precisely the polynomials in $T$. 

A modular adder is simply the unitary obtained by encoding the powers of the translation operator in parallel. The following more flexible version is appropriate for our purpose. 

\begin{definition} For integers $n,m,l\in \Z$ with $n\geq m\geq 1$, define the modular adder $M_{m,n,l}\colon H_m\otimes H_n\ra H_m\otimes H_n$ by the rule
\[ M_{m,n,l}\ket{j}\ket{k}  = \ket{j}T^{j+l}\ket{k} = \ket{j}\ket{k+j+l}  \quad j,k\in \Z.\]
\end{definition}

The modular adder is diagonalized by the Fourier transform on the second registry.   
\begin{lemma} Define $R\in\om{End}(H_n)$ to be the diagonal operator given by $R\ket{k} = e^{2\pi i k/N}\ket{k}$ for $0\leq k<N$, where $N = 2^n$. Then $T = \MC{F}^{-1}R\MC{F}$, where $\MC{F}$ is the quantum Fourier transform given by
\begin{equation}  \label{eq:QFT_def}
\MC{F}\ket{k} = 2^{-n/2}\sum_{j=0}^{N-1} e^{2\pi i jk/N}\ket{j}  \quad \text{ for } 0\leq k <N .
\end{equation} 
\end{lemma} 
\begin{proof} Let $\hat{e}_k = \MC{F}^{-1}\ket{k} = 2^{-n/2}\sum_{j=0}^{N-1}e^{-2\pi jk/N}\ket{j}$ for $0\leq k<N$. The claim is then equivalent to the observation that $T\hat{e}_k = e^{2\pi i k/N}\hat{e}_k$ for $0\leq k<N$.
\end{proof}

In view of the above result, we introduce what we coin the phase adder
\begin{equation} \label{eq:phase_adder}
P_{m,n,l} \coloneqq  (1\otimes \MC{F})M_{m,n,l}(1\otimes \MC{F}^{-1}) = \sum_{j=0}^{2^m-1} \ket{j}\bra{j}\otimes R^{j+l}.
\end{equation}
The phase adder admits a simple quantum circuit implementation that we now describe. Write $P_{m,n} = P_{m,n,0}$ and observe that $P_{m,n,l} = (1\otimes R^l)\circ P_{m,n}$. We have
\begin{equation} \label{eq:circ_R_operator}
R^l = \bigotimes_{s=0}^{n-1} P(2\pi l/ 2^{n-s}) \quad \mbox{ where } \quad P(\theta) = \left( \begin{array}{cc} 1& 0 \\ 0 & e^{i\theta} \end{array} \right) \quad \text{ for } \theta\in \R.
\end{equation} 
Let $CP_{s,t}(\theta)\colon H_m\otimes H_n\ra H_m\otimes H_n$ denote the controlled phase gate with control and target specified by $0\leq s<m$ and $0\leq t<n$. Explicitly,
\begin{equation} \label{eq:CP_def}
CP_{s,t}(\theta)\ket{j}\ket{k} = e^{i\theta j_sk_t}\ket{j}\ket{k} \quad \text{ for } 0\leq j <2^m, \; 0\leq k<2^n.
\end{equation} 

\begin{proposition} \label{prop:phase_adder}
The phase adder $P_{m,n}$ can be implemented with $O(mn)$ $2Q$-gates as follows
\[ P_{m,n} = \prod_{s = 0}^{m-1} \left(\prod_{t=0}^{n-s-1} CP_{s,t}(2\pi/2^{n-s-t})\right)  ,\]
where the product denotes gate composition. 
\end{proposition}
\begin{proof} For $0\leq j<2^m$ and $0\leq k<2^n$ we compute
\begin{align*}
 P_{m,n}&\ket{j}\ket{k} = e^{2\pi jk/2^n}\ket{j}\ket{k} = \left(\prod_{s=0}^{m-1} \prod_{t=0}^{n-1} e^{2\pi j_sk_t/2^{n-s-t}}\right)\ket{j}\ket{k} \\
 &=\left(\prod_{s=0}^{m-1} \prod_{t=0}^{n-s-1} e^{2\pi j_sk_t/2^{n-s-t}}\right)\ket{j}\ket{k}
  = \prod_{s = 0}^{m-1} \left(\prod_{t=0}^{n-s-1} CP_{s,t}(2\pi/2^{n-s-t})\right) \ket{j}\ket{k},
\end{align*}
which establishes the required formula. The number of $2Q$-gates is given by 
\[ \sum_{s=0}^{m-1} (n-s) = nm-\sum_{s=0}^{m-1}s = nm-m(m-1)/2 = O(mn).\]
\end{proof}  

\begin{remark} The $CP_{s,t}(\theta)$ gates are all diagonal in the computational basis, so the order of the factors is irrelevant.
\end{remark} 

By combining the above gate construction with \cref{eq:phase_adder}, taking into account that the quantum Fourier transform requires $O(n^2)$ $2Q$-gates, we obtain the following. 

\begin{corollary} \label{cor:complexity_adders}
Let $n\geq m \geq 1$ and $l\in \Z$. 
\begin{enumerate}[(i)]
\item The phase adder $P_{n,m,l}$ can be implemented with $O(mn)$ $2Q$-gates and $O(n)$ $1Q$-gates. 
\item The modular adder $M_{n,m,l}$ can be implemented with $O(n^2)$ $2Q$-gates and $O(n)$ $1Q$-gates. 
\end{enumerate} 
\end{corollary} 

The LCU method first introduced in \cite{Childs_Wiebe12} can be stated in the following form. 
\begin{lemma} \label{lemma:LCU}
Let $m,n\in \N$ and let $S_L,S_R\in U(H_m)$ be a pair satisfying
\[ S_L\ket{0} = \sum_{j=0}^{2^m-1} b_j\ket{j} \quad \text{ and } \quad S_R\ket{0} = \sum_{j=0}^{2^m-1}c_j\ket{j} .\]
Let $\{U_j\in U(H_n)\}_{j=0}^{2^m-1}$ be a family of unitaries and set 
$U \coloneqq \sum_{j=0}^{2^m-1} \ket{j}\bra{j}\otimes U_j$.
Then the following circuit defines a block encoding of $A \coloneqq \sum_{j=0}^{2^m-1}a_jU_j$ on the subspace $\ket{0}\otimes H_n\subset H_m\otimes H_n$, where $a_j = b_j\bar{c}_j$ for $0\leq j<2^m$. 
\[ \begin{quantikz}[wire types = {b,b}]
& \gate{S_L} & \gate[2]{U} & \gate{S_R^\dagger}&\\
& & & & 
\end{quantikz} \]
\end{lemma} 

The pair $(S_L,S_R)\in U(H_m)$ is called a state preparation pair for the vector $a = \sum_j a_j\ket{j}$. Since both $S_L\ket{0}$ and $S_R\ket{0}$ are unit vectors, it follows that
\begin{equation} \label{eq:SP_L1_bound}
||a||_1 = \sum_j |a_j| = \sum_{j} |b_j||c_j| \leq \frac12 \sum_j |b_j|^2+|c_j|^2 = 1
\end{equation}  
with equality if and only if $|b_j|=|c_j|$ for all $j$.

We can now return to our goal of block-encoding $H = i\beta D_{2p}$, where $D_{2p} = \sum_{j=1}^p\alpha_j \delta_{2j}$ is given in \cref{thm:FD_operator} and $\beta$ is a normalizing constant that will be fixed in shortly. Since
$(\Delta x)D_{2p} = (1/2)\sum_{j=1}^p (\alpha_j/j)(T^j-T^{-j})$, 
the idea is to use the modular adder $M_{m,n,-p} = \sum_{j=0}^{2^m-1}\ket{j}\bra{j}\otimes T^{j-p}$ with $m \coloneqq \lceil \log_2(2p+1)\rceil$, combined with a state preparation pair for the vector 
\begin{equation} \label{eq:SP_normalized} 
a = a(p) \coloneqq ic_p\sum_{j=1}^p \frac{\alpha_j}{2j}(\ket{p+j}-\ket{p-j}) \quad \text{ where } \quad c_p^{-1} = \sum_{j=1}^p \frac{|\alpha_j|}{j}  .
\end{equation}  
Here, the normalization ensures that $||a||_1 = 1$, which is optimal according to \cref{eq:SP_L1_bound}. Explicitly, we may take $(S_L,S_R)$ to satisfy
\begin{align} \label{eq:SP_pair_formula}
S_L\ket{0} &= i\sqrt{c_p}\sum_{j=1}^p \left(\frac{|\alpha_j|}{2j}\right)^{1/2}(\ket{p+j}+\ket{p-j}) \\
S_R\ket{0} &= \sqrt{c_p}\sum_{j=1}^{p}(-1)^{j+1}\left(\frac{|\alpha_j|}{2j}\right)^{1/2}(\ket{p+j}-\ket{p-j}). \nonumber 
\end{align}

\begin{example} For $p=3$, we compute $(\alpha_1,\alpha_2,\alpha_3) = (3/2,-3/5,1/10)$ and $c_3 = 6/11$. The vector we want to prepare is
\[ a = i\left[\frac{1}{110}(\ket{0}-\ket{6}) - \frac{9}{110}(\ket{1}-\ket{5}) - \frac{9}{22}(\ket{2}-\ket{4})\right] .\]
By the above recipe, we then seek $(S_L,S_R)$ satisfying
\begin{align*}
S_L\ket{000} &= i\left[u(\ket{010}+\ket{001}) +v(\ket{100}+\ket{101}) +w(\ket{000}+\ket{011})\right] \\
S_R\ket{000} &= \left[u(-\ket{010}+\ket{001}) +v(\ket{100}-\ket{101}) +w(-\ket{000}+\ket{011})\right],
\end{align*}
for $(u,v,w) = \left(\sqrt{\frac{9}{22}},\sqrt{\frac{9}{110}},\sqrt{\frac{1}{110}}\right)$. Set $\phi_1 = \arcsin\left(\sqrt{\frac{9}{11}}\right)$ and $\phi_2 = \arcsin\left(\sqrt{\frac{9}{10}}\right)$. Then the following circuit does the trick. 
\[ \begin{quantikz}[column sep = 0.15cm]
& \gate[3]{S_L} & \\
& \ghost{H} & \\
& \ghost{H} & \end{quantikz} 
=
 \begin{quantikz}
& & \gate{R_Y(\phi_2)} & \octrl{1} & \\
& \gate{R_Y(\phi_1)} & \octrl{-1} & \targ{} & \\
& \gate{e^{i\pi/2}} & \gate{H} & \ctrl{-1} & \end{quantikz} 
\quad
R_Y(\theta) \coloneqq \left( \begin{array}{cc} \cos(\theta) & -\sin(\theta) \\ \sin(\theta) & \cos(\theta) \end{array} \right)
\] 
The two upper rotation gates prepare $\sqrt{2}(w\ket{0}+v\ket{1}+u\ket{2})$, and from here it is clear how one arrives at the desired vector. Some minor modifications are needed to account for the signs required for $S_R$.
Replace $(\phi_1,\phi_2)$ by $(-\phi_1,\pi-\phi_2)$ and replace the two gates on the third wire by $X$ and $H$ from the left to right. 
\end{example}  

Before we continue, we include a characterization of the growth of the normalizing constant $c_p$ given in \cref{eq:SP_normalized}.

\begin{proposition} \label{prop:c_p_estimate}
The constant $c_p^{-1} = \sum_{j=1}^p |\alpha_j|/j$, where $\alpha_j = \frac{2(-1)^{j+1}(p!)^2}{(p+j)!(p-j)!}$ satisfies $c_p^{-1} = \Theta(\log{p})$. 
\end{proposition} 
\begin{proof} First, note that
\[ |\alpha_j| = \frac{2(p!)^2}{(p+j)!(p-j)!} = 2\prod_{s=1}^j \frac{p+s-j}{p+s}  = 2\prod_{s=1}^j\left(1-\frac{j}{p+s}\right).\]
Hence, $|\alpha_j| \leq 2$ for all $j$, so that $c_p^{-1} \leq 2\sum_{j=1}^p 1/j = O(\log(p))$.
To establish the lower bound, observe that for $j \leq \sqrt{p}-1$
\[ \frac12 |\alpha_j| = \prod_{s=1}^j\left(1-\frac{j}{p+s}\right) \geq \left( 1-\frac{j}{p}\right)^j \geq \left(1-\frac{\sqrt{p}}{p}\right)^{\sqrt{p}-1} \geq e^{-1}  .\]
Hence, for $r = \lfloor \sqrt{p} \rfloor -1$ $c_p^{-1} \geq \sum_{j=1}^r |\alpha_j|/j \geq 2e^{-1}
\sum_{j=1}^r 1/j = \Omega(\log(r)) = \Omega(\log(p))$ establishing the desired result. 
\end{proof} 

\begin{theorem} \label{thm:block_enc1}
Let $p\geq 1$ be an integer and set $m = \lceil \log_2(2p+1)\rceil$. Let $(S_L,S_R)$ be a state preparation pair satisfying \cref{eq:SP_pair_formula}. The quantum circuits 
\[ \begin{quantikz}[wire types = {b,b}]
& \gate{S_L} & \gate[2]{M_{m,n,-p}} & \gate{S_R^\dagger}&\\
& & & & 
\end{quantikz} 
\quad \text{ and } \quad 
\begin{quantikz}[wire types = {b,b}]
& \gate{S_L} & \gate[2]{P_{m,n,-p}} & \gate{S_R^\dagger}&\\
& & & & 
\end{quantikz} 
\]
define unitary block-encodings $U_H$ and $U_{H'}$ of the Hermitian matrices $H \coloneqq i(c_p\Delta x)D_{2p}$ and
$H' \coloneqq i(c_p\Delta x) \MC{F}^\dagger D_{2p}\MC{F}$, respectively, on the subspace $\ket{0}\otimes H_n\subset H_m\otimes H_n$, where $c_p = \Theta(\log(p))$ is defined in \cref{eq:SP_normalized}. The gate complexity is $O(p)$ for $S_L$ and $S_R$, $O(n^2)$ for $M_{m,n,-p}$ and $O(mn)$ for $P_{m,n,-p}$.   
\end{theorem}
\begin{proof} The fact that the quantum circuits defines the claimed block-encodings follows by unpacking the various definitions and applying \cref{lemma:LCU}. The gate complexity of $M_{m,n,-p}$ and $P_{m,n,-p}$ follows from \cref{cor:complexity_adders}. General state preparation on $m$ qubits has complexity $\Theta(2^m)$ in size \cite{Sun_etal21}. As $m=\lceil \log_2(2p+1) \rceil$, this leads to a gate complexity of
$O(2^m) = O(p)$ for $S_L$ and $S_R$. 
\end{proof} 

Importantly, $\om{QSVT}(U_H,\Pi,q(x)) = (I\otimes \MC{F}^\dagger)\om{QSVT}(U_{H'},\Pi,q(x))(I\otimes \MC{F})$, 
so instead of using the block-encoding of $H = ic_p\Delta x D_{2p}$ with cost $O(n^2)$, we can use the cheaper block-encoding of $H'$ with cost $O(mn)$ together with two applications of the quantum Fourier transform. This leads to the following overall resources estimate. 

\begin{lemma}\label{lem:block_enc_QSVT_complexity}
Let $U_{H'}$ be the block encoding of $H' = i(c_p\Delta x)\MC{F}^\dagger D_{2p}\MC{F}$ of \cref{thm:block_enc1} using $m+n$ qubits with $m = \lceil \om{log}_2(2p+1) \rceil$. 
Let $q(x) = q_0(x)+iq_2(x)$ be a polynomial of degree $D$ with $q_0(x),q_1(x)$ real polynomials of opposite parity satisfying $|q_0(x)|,|q_1(x)|\leq 1$ for all $x\in [-1,1]$. Then the gate complexity of applying
QSVT with the block-encoding $U_{H'}$ and the polynomial $q(x)$ is $O(Dnm)$ and the circuit uses $n+m+2$ qubits. \end{lemma} 

The proof is given in \cref{appendix:block-encoding}. 

\section{Complexity analysis}   \label{sec:complexity}
In the previous sections, we have described a quantum algorithm that can solve \cref{prob:quantum_adv_diff}. The gate complexity of the algorithm is $O(mnD)$, where $m = \lceil \log_2(2p+1)\rceil$, $n = \log_2(d/\Delta x)$ and $D$ is the degree of the polynomial approximation used in QSVT. We seek to express these parameters in terms of the input of the problem: the initial function $u_0(x)$, the evolution time $T$ and the desired precision $\eps>0$. This will be achieved in several steps.
\begin{enumerate}[(1)]
    \item We establish an error estimate for the numerical solution from which we can specify $\Delta x$ and hence $n = \log_2(d/\Delta x)$ in terms of the problem input.
    \item We determine an upper bound on the degree $D$ of the polynomial approximation needed to solve the end-to-end problem with the required precision.
    \item We assemble the pieces to obtain an expression of the gate complexity in terms of the input parameters.
\end{enumerate}
Several of the proofs in this section have been moved to \cref{appendix:complexity}.

\subsection{Numerical error estimation} \label{subsec:numerical_estimates}
In the following, we will assume that the initial function is given by a finite Fourier series
\begin{equation} \label{eq:Fourier-trunc}
u_0(x) = \sum_{k = -N_0/2}^{N_0/2-1} a_k e_k(x) ,
\end{equation} 
for some even $N_0$. 
It is well-known that a sufficiently regular $u_0$ can be approximated uniformly by a finite Fourier series as above. In fact, estimating the coefficients via the discrete Fourier transform leads to a good uniform approximation for appropriately chosen $N_0$ (see, e.g., \cite{Epstein05} for detailed results in this direction). 

The following basic lemma shows that an error made in representing the initial function propagates at worst linearly. 
\begin{lemma} \label{lem:error_propagation}
Let $u_t(x) = u(t,x)$ be a solution of $\dd_t u+ c\dd_x u = \nu \dd_x^2u$ in the domain $[0,d]$ with periodic boundary conditions.
Then 
\begin{equation}  \label{eq:heat_energy_estimate} 
||u_t||_{L^2} \leq e^{-t\nu \omega^2} ||u_0||_{L^2} + (1+e^{-t\nu \omega^2})|\rho| ,
\end{equation} 
where $\rho = d^{-1} \int_0^d u_0(x)dx$. In particular, $||u_t||_{L^2} \leq (1+2d^{-1/2})||u_0||_{L^2}$.
\end{lemma} 

As a consequence of the above, if $u_t$ and $u_t^*$ are solutions with different initial conditions, then $||u_t-u_t^*||_{L^2} \leq (1+2d^{-1})||u_0-u_0^*||_{L^2}$, showing that the error at time $t$ is controlled by the initial error. The more refined estimate in Equation \cref{eq:heat_energy_estimate} shows that the error made in the first Fourier mode is essentially preserved, while the errors made in the remaining modes dissipate with time. 

With this we take the assumption in \cref{eq:Fourier-trunc} as justified and proceed with the error estimation. Recall that we work in the spatial domain $[0,d]$ with an even number $N$ of uniformly distributed grid points. Moreover, $e_k(x) = e^{i\omega k x}$ for $k\in \Z$, where $\omega = 2\pi /d$, and the finite difference operators $\delta_{j}$, $D_{2p}$ and $D_{2p}^{(2)}$ were introduced in \cref{eq:basic_FD} and \cref{thm:FD_operator}.

For any $k\in \Z$, we have $\dd_x e_k(x) = i\omega ke_k(x)$ and $\dd_x^2 e_k(x) = -\omega^2 k^2e_k(x)$. Similarly, by inspection
\[ \delta_{2j} e_k(x) = i\frac{\sin(\omega k j\Delta x)}{j\Delta x}e_k(x) \quad \text{ and } \quad \delta_j^2 e_k(x) = -\frac{2-2\cos(\omega k j\Delta x)}{(j\Delta x)^2}e_k(x) . \]
By linearity, one obtains $D_{2p}e_k(x) = i\lambda_ke_k(x)$ and $D_{2p}^{(2)}e_k(x) = -\mu_ke_k(x)$, where
\begin{equation} \label{eq:def_lambda_mu}
\lambda_k \coloneqq \sum_{j=1}^p \alpha_j \frac{\sin(\omega k j\Delta x)}{j\Delta x} \quad \text{ and } \quad \mu_k \coloneqq \sum_{j=1}^p\alpha_j
\frac{2-2\cos(\omega k j\Delta x)}{(j\Delta x)^2}.
\end{equation}
In other words, both the differential operator $P = -c\dd_x +\nu \dd_x^2$ and the finite-difference operator $L = -cD_{2p}+\nu D_{2p}^{(2)}$ (or $L = -cD_{2p}+\nu D_{2p}^{2}$) are diagonal in the Fourier basis. In particular, if $u_0$ belongs to the subspace
\begin{equation} \label{eq:Fourier_subspace} 
\MC{T}_N \coloneqq  \om{Span}_\C\{ e_k(x) \colon -N/2 \leq k < N/2  \},
\end{equation} 
then the same apply for the exact and finite-difference solutions. Explicitly,  
\begin{equation} \label{eq:Fourier_solutions}
u(t,x) = \sum_{k = -N/2}^{N/2-1} a_ke^{-ic\omega k t-\nu \omega^2k^2 t}e_k(x) \;\; \text{ and } \;\; 
v(t,x) = \sum_{k=-N/2}^{N/2-1} a_ke^{-ic\lambda_k t-\nu \mu_kt}e_k(x) 
\end{equation}

From this it is clear that the quality of the finite-difference solution is largely controlled by the quality of the eigenvalue approximations $\lambda_k \approx \omega k$ and $\mu_k\approx \omega^2 k^2$. 

\begin{theorem} \label{thm:lambda_mu_estimates}
For a fixed positive integer $p$, let $\lambda(k)\coloneqq \lambda_k$ and $\mu(k) \coloneqq \mu_k$ be defined by the formulas in \cref{eq:def_lambda_mu}. Then the following hold true. 
\begin{enumerate}[(a)]
\item $\lambda(-k) = -\lambda(k)$ and $\mu(-k) = \mu(k)$.
\item $\lambda(k)\leq \omega k$ and $\mu(k)\leq \omega^2 k^2$ for all $k\geq 0$. 
\item The function $k\mapsto \mu(k)$ is increasing on $\in [0,N/2]$ and decreasing on $[-N/2,0]$. The function $\lambda(k)$ is positive on $[0,N/2]$ and negative on $[-N/2,0]$. 
\item We have the following estimates
\begin{align*}
&|\omega k-\lambda_k| \leq \frac{(p!)^2}{(2p+1)!} |\omega k|^{2p+1} (\Delta x)^{2p}  \\  
& |\omega^2k^2-\lambda_k^2| \leq \frac{2(p!)^2}{(2p+1)!}|\omega k|^{2p+2}(\Delta x)^{2p} \\
& |\omega^2k^2-\mu_k|\leq \frac{2(p!)^2}{(2p+2)!} |\omega k|^{2p+2}(\Delta x)^{2p}  .
\end{align*}
\end{enumerate}
\end{theorem} 

We can freely pass between functions defined on the grid and functions in $\MC{T}_N$ by the following result.  

\begin{lemma} \label{lemma:norm_comparison}
For an even positive integer $N$, let $\MC{T}_N$ be defined as in \cref{eq:Fourier_subspace} and equip it with the 
$L^2$ inner product $(f,g) = \int_0^d f(x)\overline{g(x)}dx$. Then the evaluation map $\om{ev}\colon \MC{T}_N\ra \C^N$ given by $\om{ev}(q)_j = q(j\Delta x)$, $0\leq j<N$, where $\Delta x = d/N$, is an isomorphism and $||q||_{L^2} = (\Delta x)^{1/2} ||\om{ev}q||$ for all $q$ when $\C^N$ is equipped with the standard Hermitian inner product. 
\end{lemma}

The error in the finite-difference solution is characterized as follows.  

\begin{theorem} \label{thm:L2_estimate}
Fix a positive integer $p$. Let $u_t(x) = u(t,x)$ and $v_t(x) = v(t,x)$, as given in \cref{eq:Fourier_solutions}, be
solutions of the equations
\[ \dd_tu(t,x) = (-c\dd_x+\nu \dd_x^2)u(t,x) \quad \text{ and } \quad \dd_tv(t,x) = (-cD_{2p}+\nu D_{2p}^{(2)})v(t,x), \]
respectively, with the same initial condition. Then
\begin{equation} \label{eq:diff_difference_estimate}
 ||u_t-v_t||_{L^2} \leq te^{-\nu t\mu_1}  (\Delta x)^{2p} \left[ c^2C_p^2 ||u_0^{(2p+1)}||_{L^2}^2 +  \nu^2(C'_p)^2||u_0^{(2p+2)}||_{L^2}^2 \right]^{1/2},    
\end{equation}
where $u_t(x) = u(t,x)$, $v_t(x) = v(t,x)$, $C_p = \frac{(p!)^2}{(2p+1)!}$ and $C'_p = \frac{2(p!)^2}{(2p+2)!}$.
\end{theorem}

\begin{remark} The same result is true with $D_{2p}^{(2)}$ replaced by $D_{2p}^2$, provided we
make the following modifications: Replace $\mu_k$ by $\lambda_k^2$ in the definition of $v(t,x)$, replace $\mu_1$ by $\lambda_1^2$ in the main estimate, and replace the constant $C_p'$ in the main estimate by $C_p'' = \frac{2(p!)^2}{(2p+1)!}$ (compare \cref{thm:lambda_mu_estimates} part (d)).
\end{remark}

In our complexity analysis, we will work with a slightly weaker form of the bound in \cref{eq:diff_difference_estimate}. First, we will assume that $u_0$ can be expressed with $N_0$ Fourier modes and work with $N\geq N_0$ grid points. Having invoked this lower bound, we can avoid the $\Delta x$-dependence of $\mu_1 = \mu_1(\Delta x,p)$ by substituting the smaller quantity $\mu_1' \coloneqq \mu_1((\Delta x)_0,p)$, where $(\Delta x)_0 = d/N_0$. The formula we will use is the following 
\begin{equation} \label{eq:simplified_bound}
||u_t-v_t||_{L^2} \leq te^{-\nu t\mu_1'}(\Delta x)^{2p} \left[c||u_0^{(2p+1)}||_{L^2}+ \nu ||u_0^{(2p+2)}||_{L^2} \right]. 
\end{equation} 

\subsection{Polynomial approximations}
The goal is to produce a polynomial approximation of the function  (recall \cref{eq:exp_QSVT})
\begin{equation}  \label{eq:exp_approx_func}
 f(x;M_1,M_2) = e^{-M_1x^2 + M_2ix} \quad \text{ for } \quad M_1,M_2\in \R_{\geq 0}    .
\end{equation}
More precisely, given $M_1$, $M_2$ and $\eps>0$, we seek a polynomial $q(x)$ of minimal degree satisfying 
\[ ||f(; M_1,M_2)-q(x)||_{\infty,[-1,1]} \coloneqq \sup_{x\in [-1,1]}|f(x;M_1,M_2)-q(x)| <\eps  .\]
It is known that for analytic functions with absolutely summable Chebyshev coefficients, the truncation of the Chebyshev expansion is close to the optimal polynomial approximation of a fixed degree (see \cite[Prop.~2.2]{Aggarwal_Alman22}). This strategy has been carried out for $f(x;M_1,0)$ in \cite{QSVT19} and $f(x;0,M_2)$ in \cite{Low_Chuang17, Aggarwal_Alman22}. We combine these results to construct a polynomial approximation of the product $f(x;M_1,M_2) = f(x;M_1,0)f(x;0,M_2)$ in this section. This has the advantage of producing explicit expressions for the polynomials in the Chebyshev basis, which can then be used to accurately compute the corresponding angle sequences with the method of \cite{DongMeng21}. 

For $z\in \C$ and $\theta \in \R$ one has the Jacobi-Anger expansion
\begin{equation} \label{eq:ja_expansion}
e^{iz\cos(\theta)} = J_0(z) + 2\sum_{n=1}^\infty i^n J_n(z)\cos(n\theta),
\end{equation}
where $J_n(z)$ is the $n$'th Bessel function of the first kind \cite{HandbookFunc}. This leads to the Chebyshev expansion
\begin{equation} \label{eq:ja_cheb_expansion}
e^{izx} = J_0(z) + 2\sum_{n=1}^\infty i^n J_n(z)T_n(x), \quad x\in [-1,1],
\end{equation}
where $T_n(x)$ is the $n$'th Chebyshev polynomial of the first kind specified by the relation $T_n(\cos(\theta)) = \cos(n\theta)$. By equating real and imaginary parts, we obtain the Chebyshev expansions of $\cos(Mx)$ and $\sin(Mx)$. Introduce the corresponding truncations 
\begin{align} \label{eq:ja_trunc}
 C_R(x;M) &= J_0(M) + 2\sum_{k=1}^R(-1)^k J_{2k}(M)T_{2k}(x) \\
S_R(x;M) &= 2\sum_{k=0}^R (-1)^k J_{2k+1}(M)T_{2k+1}(x). \nonumber
\end{align}

The following can be extracted from \cite[Lem.~57,~59]{QSVT19}.  

\begin{lemma} \label{lem:trig_trunc}
Given $M>0$ and $\eps \in (0,1)$, the minimal $R$ for which 
\[ |C_R(x;M)-\cos(Mx)|\leq \eps \quad \text{ and } \quad |S_R(x;M)-\sin(Mx)|\leq \eps  \]
for all $x\in [-1,1]$, satisfies $R = O(\max\left( M, \frac{\log(\eps^{-1})}{\log(e+M^{-1}\log(\eps^{-1}))}\right)$.
\end{lemma}

The Chebyshev expansion of the Gaussian can be derived by substituting $\theta \mapsto  2\theta$ and $z\mapsto iM/2$ in the Jacobi-Anger expansion \cref{eq:ja_expansion}.
\begin{equation}\label{eq:2exp_cheb} 
e^{-Mx^2}  = \hat{I}(M/2)+2\sum_{n=1}^\infty (-1)^n\hat{I}_n(M/2)T_{2n}(x) ,
\end{equation} 
where $\hat{I}_n(y) = e^{-y}i^{-n}J_n(iy)$ is the scaled modified Bessel function. Define the corresponding truncation by
\begin{equation} \label{eq:}
E_R(x;M) \coloneqq \hat{I}_0(M/2) + 2\sum_{k=1}^R (-1)^k \hat{I}_k(M/2)T_{2k}(x). 
\end{equation} 

The following is a simple improvement of \cite[Cor.~3]{Low_Chuang17} that can be deduced from \cite[Thm.~2.1]{Aggarwal_Alman22}. 

\begin{lemma} \label{lem:exp_trunc} 
Given $M>0$ and $\eps\in (0,1)$, the minimal degree $R$ for which 
\[ |E_R(x;M) - e^{-Mx^2}|\leq \eps \quad \text{ for all } x\in [-1,1] , \]
satisfies $R = O(\max \left(\sqrt{ M\log(\eps^{-1})}, \frac{\log(\eps^{-1})}{\log(e+M^{-1}\log(\eps^{-1}))}\right) )$.
\end{lemma} 

We now combine the above results to produce the desired polynomial approximation. Note that we use the slightly weaker bound $\log(\eps^{-1}) \geq \frac{\log(\eps^{-1})}{\log(e+M^{-1}\log(\eps^{-1}))}$. This is justified because the right-hand bound is primarily relevant when $M = o(\log(\eps^{-1})$ (see \cite{Aggarwal_Alman22}), which has little interest in our applications since $M_1$ and $M_2$ contain a factor of the form $\eps^{-s}$ for some $s\in (0,1)$.  

\begin{proposition} \label{prop:product_approx} 
Given $M_1,M_2\geq 0$ and $\eps>0$, there are polynomials $P_e(x)$ and $P_o(x)$ of degree $2R$ and $2R+1$, respectively, that are bounded by $1$ on $[-1,1]$ and satisfy 
\[ |e^{-M_1x^2}\cos(M_2x) - P_e(x)| \leq \eps \quad \text{ and } \quad
|e^{-M_1x^2}\sin(M_2x) - P_o(x)|\leq \eps    \]
for all $x\in [-1,1]$, where $R = O(\max(\sqrt{M_1\log(\eps^{-1})}+M_2, \log(\eps^{-1}))$. 
\end{proposition} 
\begin{proof} In the following, let $||\cdot || = ||\cdot ||_{\infty,[-1,1]}$. We will rely on the following basic product bound. If $||f-p||<\eps/2$, $||g-q||\leq \eps/2$ and $||f||,||g||,||p||,||q||\leq 1$, then 
\begin{equation} \label{eq:proof_product_bound}
||fg-pq||\leq ||f||\cdot ||g-q|| + ||q||\cdot ||f-p|| \leq \eps  .
\end{equation} 
In our situation, $f(x) = e^{-M_1x^2}$ and $g(x) = \cos(M_2x)$ or $g(x) = \sin(M_2x)$. We will restrict ourselves to the case $g(x) = \cos(M_2x)$ as the other case is entirely analogous. By \cref{lem:exp_trunc,lem:trig_trunc}, we can find $R_1$ and $R_2$ satisfying
\begin{equation}  \label{eq:R12_bounds} 
R_1 = O(\max( \sqrt{M_1\log(\eps^{-1})},\log(\eps^{-1})) \quad \text{ and  } \quad 
R_2 = O(\max(M_2,\log(\eps^{-1})).
\end{equation} 
such that $||e^{-M_1x^2}-E_{R_1}(x;M_1)||\leq \eps/4$ and $||\cos(M_2x)-C_{R_2}(x;M_2)||\leq \eps/4$. Then with $p(x) = (1+\eps/4)^{-1}E_{R_1}(x;M_1)$ and $q(x)= (1+\eps/4)^{-1}C_{R_2}(x;M_2)$, we have 
$||f-p||\leq \eps/2$, $||g-q||\leq \eps/2$ and $||p||,||q||\leq 1$ due to the normalization. The product bound \cref{eq:proof_product_bound}, then shows that the degree $2R = 2(R_1+R_2)$ polynomial $P_e(x) = p(x)q(x)$ does the trick. By our choice of $R_1$ and $R_2$ we have $R = O(\max(\sqrt{M_1\log(\eps^{-1})}+M_2, \log(\eps^{-1}))$ as claimed. 
\end{proof} 

\subsection{Gate complexity of the algorithm}
Fix $N_0>0$ and assume that $u_0(x)$ takes the form given in \cref{eq:Fourier-trunc}. In particular, $u_0(x)$ belongs to the space $\MC{T}_N$ (see \cref{eq:Fourier_subspace}) for any $N\geq N_0$. At this point, we have the following three (approximate) solutions. 
\begin{enumerate}[(1)]
    \item The exact solution $u_t \in \MC{T}_N$.
    \item The finite-difference solution $v_t = e^{Lt}u_0\in \MC{T}_N$, where $L = -cD_{2p}+\nu D_{2p}^2$.
    \item The quantum solution $w_t = P(H)u_0 \in \MC{T}_N$, where $H = ic_p \Delta x D_{2p}$ and $P(x) \approx f(x;M_1,M_2)$ with $M_1 = \frac{\nu t}{(c_p\Delta x)^2}$ and $M_2 = \frac{ct}{c_p\Delta x}$. 
\end{enumerate}
The exact solution is uniquely defined for all $N\geq N_0$ and depends only on $c,\nu\geq 0$ and the evolution time $t$. The finite-difference solution depends on $D_{2p}$, which is specified by $p\geq 1$ and $\Delta x = d/N$, and the quantum solution $w_t$ depends in addition on the choice of polynomial approximation $P$. Recall from \cref{lemma:norm_comparison} that the evaluation map $\om{ev}\colon \MC{T}\ra \C^N$ is an isomorphism that preserves the inner products up to a scaling factor. Write $\bar{u}_t = \om{ev}(u_t)$ and similarly for $v_t$ and $w_t$. The objective of \cref{prob:quantum_adv_diff} is to arrange that 
\begin{equation}  \label{eq:goal_bound}
||\bar{u}_T - \bar{w}_T || \leq \eps ||\bar{u}_0|| \; \Leftrightarrow \;  
||u_T - w_T ||_{L^2} \leq \eps ||u_0||_{L^2}  
\end{equation} 
In other words, since the required bound is relative, we can just as well work in $\MC{T}_N$ with the $L^2$ inner product.  

\begin{theorem} \label{thm:main_complexity}
Let the equation parameters $c,\nu,\geq 0$, the initial function $u_0(x)$, the evolution time $T>0$ and the precision $\eps>0$ be given. Furthermore, assume that $u_0(x)$ can be expressed exactly as $N_0$ Fourier modes as in \cref{eq:Fourier-trunc}. Given $p\geq 1$, set 
\begin{align}  \label{eq:thm_n_formula}
& n = \max( \lceil \log_2 (d [2\tau B\eps^{-1}||u_0||^{-1}_{L^2}]^\frac{1}{2p} )\rceil,\lceil \log_2N_0 \rceil), \quad \text{ where } \\  
B = B(c,\nu,& u_0 ,p) = c||u_0^{(2p+1)}||_{L^2}+ \nu ||u_0^{(2p+2)}||_{L^2}
\; \text{ and } \;
\tau = \tau(T,\nu,N_0) = Te^{-\nu T \mu_1'} \nonumber 
\end{align} 
and let $\Delta x = d/2^n$. Then our quantum algorithm utilizing the block-encoding of $D_{2p}$ can solve \cref{prob:quantum_adv_diff} using $n + \lceil \log_2(2p+1) \rceil + 2$ qubits with a gate complexity of
\begin{equation}  \label{eq:general_complexity} 
O(\max[(\Delta x)^{-1}(\sqrt{\nu T\log(\eps^{-1})}+cT)\log(p),\log(\eps^{-1})]\log(d(\Delta x)^{-1})\log(p))
\end{equation} 
\end{theorem}
\begin{proof} We seek to fix $n$ and hence $\Delta x = d/2^n$ such that $||u_T-w_T||\leq \eps ||u_0||_{L^2}$. 
Observe that by the triangle inequality, we have 
\begin{align}
||u_T-w_T||_{L^2} &\leq ||u_T - v_T||_{L^2} + ||v_T-w_T||_{L^2} \\
&= ||u_T-v_T||_{L^2} + ||(f(H;M_1,M_2)-P(H))u_0||_{L^2}  \nonumber  \\
&\leq  ||u_T-v_T||_{L^2} + ||f(x;M_1,M_2)-P(x)||_{\infty,[-1,1]} ||u_0||_{L^2}. \nonumber 
\end{align} 
The latter quantity is bounded by $\eps$ as long as we can arrange
\begin{equation}   \label{pf:goal_bound}
||u_T-v_T||_{L^2} \leq \eps||u_0||_{L^2}/2 \quad \text{ and } ||f(x;M_1,M_2)-P(x)||_{\infty,[-1,1]} \leq \eps/2. 
\end{equation} 
First, we have $||u_T-v_T||_{L^2} \leq \tau B (\Delta x)^{2p}$ by \cref{thm:L2_estimate} or more precisely \cref{eq:simplified_bound}. Hence, for $\Delta x\leq (\eps ||u_0||_{L^2}/(2\tau B)^{1/(2p)}$, we obtain $||u_T-v_T||\leq \eps||u_0|_{L^2}/2$. Since we seek $\Delta x$ in the form $d/2^n$, the optimal choice is 
\[ n = \max( \lceil \log_2 [d(2\tau B/(\eps||u_0||))^{1/(2p)}] \rceil, \lceil \log_2 N_0\rceil) \]
as in \cref{eq:thm_n_formula}. 

Next, according to \cref{prop:product_approx} we can find a polynomial $P(x)$ satisfying the second inequality in \cref{pf:goal_bound} of degree 
$D  = O(\max(\sqrt{M_1\log(\eps^{-1})}+M_2, \log(\eps^{-1}))$. As $M_1 = \nu T/(c_p \Delta x)^2$ and $M_2 = cT/(c_p \Delta x)$, we obtain 
\begin{equation}  \label{pf:degree_formula}  
D = O(\max( (\Delta x)^{-1}log(p)(\sqrt{\nu T\log(\eps^{-1})}+ cT),\log(\eps^{-1})),
\end{equation} 
where we have used $c_p^{-1} = \Theta(p)$ by \cref{prop:c_p_estimate}. 
According to \cref{lem:block_enc_QSVT_complexity}, the quantum algorithm uses $n+m+2$ qubits with $m = \lceil \log_2(2p+1) \rceil$ and has gate complexity $O(Dnm)$. In view of \cref{pf:degree_formula} and the formula
$n = \log_2(d(\Delta x)^{-1})$, we arrive at the stated gate complexity. 
\end{proof}

We will now make the complexity bound more explicit by imposing a few conditions that are satisfied in most situations of interest. 

\begin{corollary}  \label{cor:complexity_simplified}
In the situation of \cref{thm:main_complexity}, assume that 
\begin{equation} \label{eq:n_dominates_N0}
n = \lceil \log_2(d(2\tau B||u_0||^{-1}_{L^2}\eps^{-1})^{\frac{1}{2p}}\rceil \geq \lceil \log_2 N_0 \rceil. 
\end{equation}
and that we confine the parameters $c$, $\nu$, $T$ and $u_0$ so that at least one of the quantities
$cT(\tau B||u_0||^{-1}_{L^2})^{\frac{1}{2p}}$ or $\nu T(\tau B||u_0||_{L^2}^{-1})^{\frac{1}{2p}}$ is bounded from below. Then the gate
complexity is given by
\begin{equation} 
\widetilde{O}\left( \left(\sqrt{\nu T\log(\eps^{-1})}+cT\right)\left[ Te^{-\nu T\mu_1'} \left(\frac{ c||u_0^{(2p+1)}||_{L^2} + \nu ||u_0^{(2p+2)}||_{L^2}}{\eps||u_0||_{L^2}}\right) \right]^{\frac{1}{2p}}\log(p)^2 \right.)
\end{equation} 
\end{corollary}
\begin{proof} Let $\bar{B}\coloneqq B/||u_0||_{L^2}$ and observe that assumption \cref{eq:n_dominates_N0} implies that $(2\tau \bar{B}/\eps)^{\frac{1}{2p}} \leq (\Delta x)^{-1} \leq 2(2\tau \bar{B}/\eps)^{\frac{1}{2p}}$. Note that $\log(\eps^{-1}) \leq 2pe^{-1} \eps^{-\frac{1}{2p}}$ for all $\eps>0$ by basic calculus. Therefore, if we restrict the parameters $c$, $\nu$, $T$ and $u_0$ so that $(\tau \bar{B})^{\frac{1}{2p}}(\sqrt{\nu T\log(\eps^{-1})}+cT)$ is bounded below, then we can replace the maximum in \cref{eq:general_complexity}
by the first argument. It is not hard to see that it suffices to require one of the quantities $cT(\tau \bar{B})^{\frac{1}{2p}}$ or $\nu T(\tau \bar{B})^{\frac{1}{2p}}$ to be bounded below for this to be true. The complexity bound in \cref{eq:general_complexity} now simplifies to 
\[ \widetilde{O}((\Delta x)^{-1}(\sqrt{\nu T\log(\eps^{-1})}+cT)\log(p)) ,\]
where the $\widetilde{O}$ notation hides the logarithmic factor $\log(d(\Delta x)^{-1})$.
If we substitute $(\Delta x)^{-1} = O( (\tau \bar{B}\eps^{-1})^{\frac{1}{2p}}$, we arrive at the stated formula. 
\end{proof} 

\section{The multidimensional case}  \label{sec:high-dim_case}
Consider the general advection-diffusion equation $\dd_t u + c\cdot \nabla u = \nu \Delta u$,
where $\nabla u = (\dd_{x_1}u,\dd_{x_2}u,\ldots, \dd_{x_m}u)$ is the gradient and $\Delta u = \sum_{j=1}^m \dd_{x_j}^2u$ is the Laplacian. Observe that we have splitting
\[ P = -c\cdot \nabla + \nu \Delta  = \sum_{j=1}^m -c_j \dd_{x_j} + \nu \dd_{x_j}^2 \eqqcolon \sum_{j=1}^m P_j  .\]
If we take a uniform grid with $N$ points in each coordinate direction, the resulting discrete space takes the form $\R^{N^m} = (\R^N)^{\otimes m}$. Introduce the finite-difference approximation $L = \sum_{j=1}^m I^{\otimes(j-1)}\otimes L_j \otimes I^{\otimes(d-j)}$, where $L_j \approx P_j$ is one of the finite-difference approximations in \cref{eq:FD_approx}.
The associated difference-differential equation becomes $V'(t) = LV(t)$ with the solution
$V(t) = \bigotimes_{j=1}^m e^{L_jt}$. The $m$-dimensional approximation can therefore be implemented with $m$ instances of the $1$-dimensional algorithm in parallel, which corresponds to taking the tensor product of the resulting block-encodings. This increases the ancilla count by a factor of $m$, but preserves the gate depth. Alternatively, one can compose the block-encodings with only $\lceil log_2(m)\rceil$ additional ancillas, but this multiplies the gate depth by $m$ (see \cite[Sec.~10.2]{QAlg25}). In our two-dimensional implementation we use the composition method $e^{Lt} = (e^{L_1t}\otimes I)\circ (I\otimes e^{L_2t})$. 

\section{Numerical results}  \label{sec:numerical_results}
In this section, we display numerical results with a computational implementation of the quantum algorithm for finite-difference operators of orders $2p = 2,4,6,14$ proposed in this paper for a selection of initial conditions, times, and parameters. The code is written in qiskit \cite{qiskit24} and is available in an accompanying \href{https://github.com/GOHelle/QC_advection-diffusion_sim}{Github repository} \cite{gitrep_GOH}. The quantum circuits are simulated with the freely available aer simulator. To display the internal accuracy of the algorithm, we have used exact statevector simulation, and compared the results to a high accuracy Fourier approximation ($\text{error} \sim 10^{-8}$ using the error formula below \cref{eq:error_measure}) of the exact solution given in \cref{eq:exact_solution}. In the case of pure advection, the exact solution $u(t,x-ct)$ is used instead. The error displayed is computed as 
\begin{equation} \label{eq:error_measure}
\text{error}  = \max_{0\leq j<2^n} ||v_T(j\Delta x)-u_T(j\Delta x) ||,     
\end{equation} 
where $v_T$ is the solution computed by the quantum algorithm after rescaling and $u_T$ is the approximate exact solution. Note that the cost of preparing the initial state, a normalized version of $u_0(x)$ evaluated on the grid, is not included in the reported gate counts.

We are particularly interested in comparing the numerical behaviour at different orders, where a finer resolution, that is, additional spatial qubits, are used at lower order to achieve a similar accuracy to the simulation at higher order. We use the spatial domain $[0,4]$ throughout. Therefore, $n$ spatial qubits correspond to $2^n$ grid points and a step size of $\Delta x = 2^{2-n}$. The resources used in the runs are quantified in terms of gate counts; $\om{CNOT}$-gates and $1$-qubit gates, the total number of qubits and the accuracy of the output. 

Since our algorithm depends on block-encoding, one has to post-select on the ancilla register being in the $\ket{0}$-state. This is called the success probability. In an idealized implementation, the success probability is $||\bar{u}_T||/||\bar{u}_0||$. In our implementation, an additional factor $1/2$ is introduced
when QSVT is applied with non-definite polynomials, as detailed in the supplementary material, which occurs when advection is present. In addition, a factor $0.95$ is introduced for numerical stability when we compute the angle sequence corresponding to a polynomial approximation. This leads to a reduced success probability of $(0.95\times 0.5)^2 \approx 0.2256$ in the $1d$-simulations and $(0.95\times 0.5)^4 \approx 0.0509$ in the $2d$ simulations, since we compose two instances of the one-dimensional algorithm. 

\subsection{1D tests}

\begin{figure}[htbp]
\centering 
\includegraphics[scale=0.53]{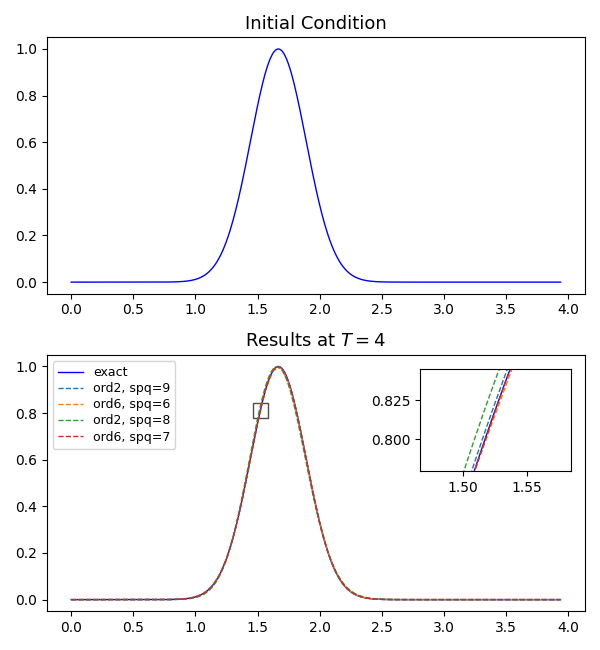}
\caption{Comparison of methods of order $2$ and $6$ and exact solution (bottom panel) for the QSVT-based solution of the 1d advection equation with speed $c=1$ and Gaussian initial conditions $u_0(x) = \exp(-10(x-5/3)^2)$ (top panel). The number of spatial qubits used is denoted by spq.}
\label{fig:figure1}
\end{figure}

\paragraph{Gaussian wave}
In a first one-dimensional test, we consider the case of pure advection ($\nu = 0$) with speed $c=1$ and Gaussian initial data $u_0(x) = e^{-10(x-5/3)^2}$ for $x\in[0,4]$ that is run until time $T = 4$, corresponding to a full period (\cref{fig:figure1} and \cref{tab:table1}).

We compare the order $2$ method with $8$ and $9$ spatial qubits to the order $6$ method with $6$ and $7$ spatial qubits (bottom panel in \cref{fig:figure1}). The numerical results are visually indistinguishable from the exact solution, and this is reflected in the error values displayed in \cref{tab:table1}. The order $6$ method outperforms the order $2$ method on all parameters: accuracy, gate counts and total qubit count. Indeed, both the run at order $2$ with $9$ spatial qubits and the run at order $6$ with $6$ spatial qubits have an error $\sim 10^{-3}$, but the order $2$ method uses $\sim 2.5$ times as many $\om{CNOT}$ and $1$-qubit gates.

It is also interesting to see how the gate count and accuracy scale as the number of spatial qubits is increased by one. Note that this corresponds to replacing the step size $\Delta x$ by $\Delta x/2$. For the order $2$ method, the accuracy improves by around a factor of ten, while the gate count roughly doubles. For the order $6$ method the accuracy improves by a factor of $10^{-2}$, while the gate count scales slightly better than in the second order method.

\begin{table}[htbp]
\centering
\footnotesize
\caption{Data table for the 1D Gaussian wave simulation in \cref{fig:figure1}.}
\begin{tabularx}{0.75\columnwidth}{cccXXcc}
\toprule
Order & Spatial  & Total  & Error & Success  & 1-qubit  & CNOT  \\
 &  qubits &  qubits & (sv) &  rate &  gates &  gates \\ \midrule
2 & 8 & 12  & 2.042e-02  & 0.2256 & 23433 & 16150 \\ \midrule
2 & 9 & 13  & 5.047e-03  & 0.2256 & 49298 & 33889 \\ \midrule
6 & 6 & 11  & 1.856e-03  & 0.2256 & 18658 & 13636 \\ \midrule
6 & 7 & 12  & 3.298e-05  & 0.2256 & 37386 & 27130 \\ \bottomrule
\end{tabularx}

\label{tab:table1}
\end{table}

\begin{figure}[htbp] 
\centering
\includegraphics[scale=0.53]{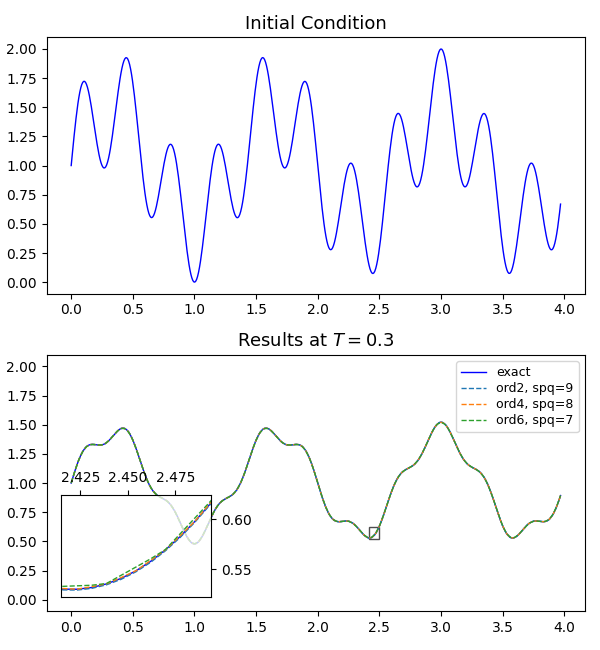}
\caption{Comparison of QSVT-based methods of order $2$, $4$ and $6$ and exact solution (bottom panel) for the pure diffusion equation with $\nu = 0.02$ and initial condition $u_0(x) = 1+(1/2)[\sin(3\pi x/2)+\sin(11\pi x/2)]$ (top panel). The number of spatial qubits used is denoted by spq.}
\label{fig:figure2}
\end{figure}

\paragraph{Sum of sine waves}
Next, we consider the pure diffusion equation ($c=0$) with $\nu = 0.02$, and initial data 
$u_0(x) = 1+(1/2)[\sin(3\pi x/2)+\sin(11\pi x/2)]$ for $x\in [0,4]$ evolved until time $T = 0.3$
(\cref{fig:figure2} and \cref{tab:table2}).

\begin{table}[htbp]
\centering
\footnotesize
\caption{Data table for the sum of sine waves simulation in \cref{fig:figure2}.}
\begin{tabularx}{0.75\columnwidth}{cccXXcc}
\toprule
Order & Spatial  & Total  & Error & Success  & 1-qubit  & CNOT  \\
 &  qubits &  qubits & (sv) &  rate &  gates &  gates \\ \midrule
2 & 9 & 12  & 9.362e-04  & 0.7937 & 10884 & 7686 \\ \midrule
4 & 8 & 12  & 5.256e-05  & 0.7937 & 10069 & 7459 \\ \midrule
6 & 7 & 11  & 4.998e-05  & 0.7937 & 5378 & 3054 \\ \midrule
\end{tabularx}
\label{tab:table2}
\end{table}

We compare the methods of order $2$, $4$ and $6$, with $9$, $8$ and $7$ spatial qubits, respectively. The error is $\sim 10^{-5}$ in each case, and the numerical solutions are visually indistinguishable from the exact solution in \cref{fig:figure2}. The methods of order $2$ and $4$ have almost identical total qubit counts and gate counts. In contrast, the order $6$ method uses one qubit less and about half as many gates. The methods of order $4$ and $6$ both uses $3$ ancillas (the difference between the total number and the number of spatial qubits in \cref{tab:table2}), while the method of order $2$ uses only $2$. This explains why the step from order $4$ to order $6$ has a more drastic effect than the step from order $2$ to $4$. In fact, the highest order we can achieve with $3$ ancillas is $6$.

\begin{figure}[htbp] 
\centering
\includegraphics[scale=0.53]{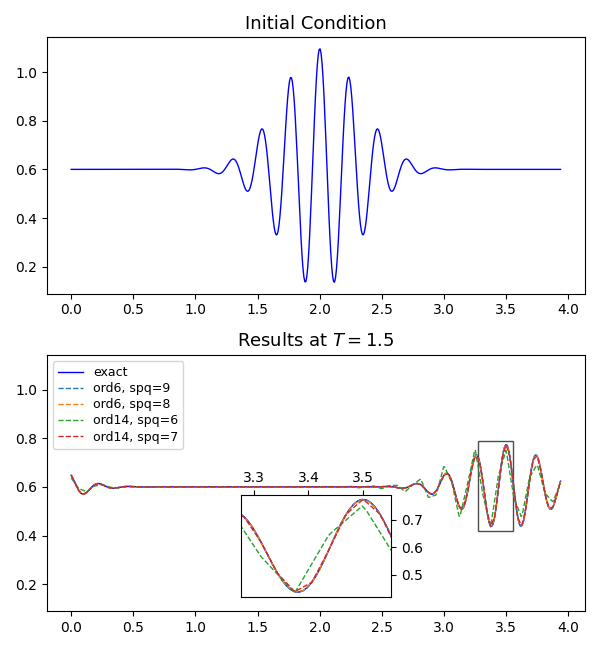}
\caption{Comparison of QSVT-based methods of order $6$ and $14$ and exact solution (bottom panel) for the advection-diffusion equation with $c = 1$, $\nu = 10^{-3}$ and initial data a wave packet (top panel, \cref{eq:wavepack})}
\label{fig:figure3}
\end{figure}

\paragraph{Wave packet}
Next, we consider a wave packet given by 
\begin{equation} \label{eq:wavepack}
u_0(x) = \frac{3}{5}+ \frac{1}{2} e^{-5(x-2)^2}\cos\left(\frac{17\pi}{2} (x-2)\right)
\end{equation}
subject to advection with $c=1$ and mild diffusion with $\nu = 10^{-3}$ for $x\in[0,4]$, evolved until time $T = 1.5$ (\cref{fig:figure3} and \cref{tab:table3}). 

We compare the order $6$ method with $8$ and $9$ spatial qubits, respectively, to the order $14$ method with $6$ and $7$ qubits, respectively. In contrast to the previous runs, the order $6$ method outperforms the order $14$ method in accuracy. With $8$ spatial qubits for the order $6$ method and $6$ spatial qubits for the order $14$ method, the gate counts are slightly in favor of the higher order method, but its accuracy is worse by two orders of magnitude. With one additional spatial qubit for both methods, the lower order method still has superior accuracy, albeit with a narrower gap. The poor performance of the order 14 method can be explained by the fact that the initial function needs more than $6$ spatial qubits to be properly resolved. Furthermore, 
the higher order methods are only expected to be favorable for resolutions under a certain threshold due to larger prefactors. 

\begin{table}[htbp]
\centering
\footnotesize
\caption{Data table for the wave pack simulation in \cref{fig:figure3}.}
\begin{tabularx}{0.75\columnwidth}{cccXXcc}
\toprule
Order & Spatial  & Total  & Error & Success  & 1-qubit  & CNOT  \\
 &  qubits &  qubits & (sv) &  rate &  gates &  gates \\ \midrule
6 & 8 & 13  & 2.662e-04  & 0.2398 & 27135 & 19578 \\ \midrule
6 & 9 & 14  & 4.334e-06  & 0.2398 & 55333 & 39729 \\ \midrule
14 & 6 & 12  & 5.429e-02  & 0.2399 & 21663 & 17331 \\ \midrule
14 & 7 & 13  & 1.483e-05  & 0.2398 & 40290 & 32029 \\ \midrule
\end{tabularx}
\label{tab:table3}
\end{table}

\subsection{2D tests}
\begin{figure}[htbp] 
\centering

\includegraphics[scale=0.56]{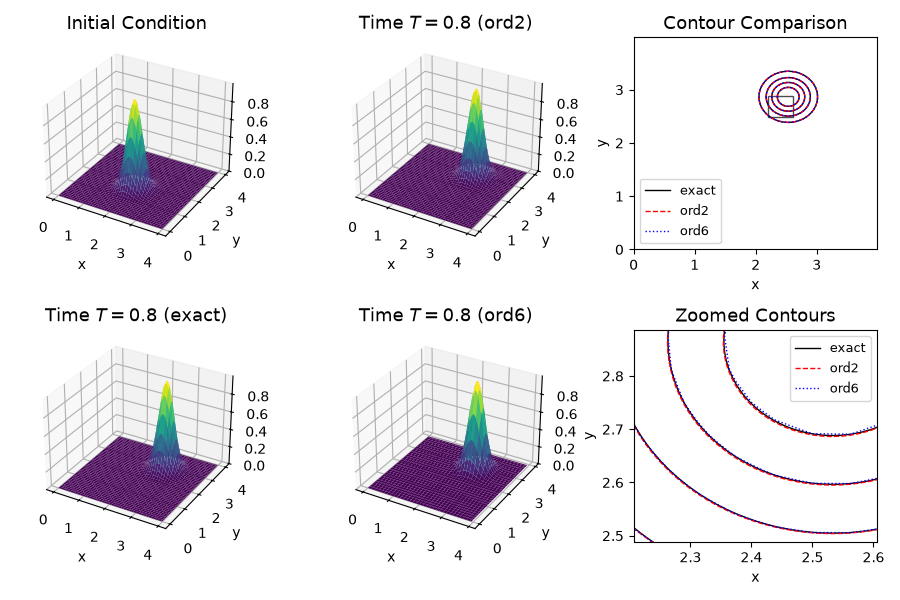}        
\caption{Comparison of QSVT-based methods of order $2$ and $6$ (bottom panels) and exact solution (top right panel) for the 2D advection equation with $c = (3/2,2/3)$ and Gaussian initial data (top left panel, \cref{eq:gaussian_2d}).}  
\label{fig:figure5}
\end{figure}

\paragraph{Two-dimensional Gaussian}
Here we consider pure advection with velocity $c = (3/2,2/3)$ and initial data
\begin{equation} \label{eq:gaussian_2d}
u_0(x,y) = \exp(-7(x-5/3)^2 - 7(y-2)^2)  
\end{equation}
for $(x,y)\in[0,4]^2$ evolved until time $T = 0.8$ (\cref{fig:figure5} and \cref{tab:table5}). 

We compare the order $2$ method with $16$ spatial qubits ($8$ in each coordinate direction) to the order $6$ method with $12$ spatial qubits. In this case, the order $6$ method outperforms the order $2$ method on all metrics by a relatively small amount. As the evolution time increases, the distinction is expected to be more prominent as observed in the 1d Gaussian wave simulation. This is also compatible with the complexity estimate in \cref{eq:general_complexity}.

\begin{table}[htbp]
\centering
\footnotesize
\caption{Data table for the two-dimensional Gaussian simulation in \cref{fig:figure5}.}
\begin{tabularx}{0.75\columnwidth}{cccXXcc}
\toprule
Order & Spatial  & Total  & Error & Success  & 1-qubit  & CNOT  \\
 &  qubits &  qubits &  &  rate &  gates &  gates \\ \midrule
2 & 16 & 21  & 4.197e-03  & 0.0509 & 14546 & 9630 \\ \midrule
6 & 12 & 18  & 2.265e-04  & 0.0509 & 13183 & 9176 \\ \midrule
\end{tabularx}
\label{tab:table5}
\end{table}

\begin{figure}[htbp] 
\includegraphics[scale=0.56]{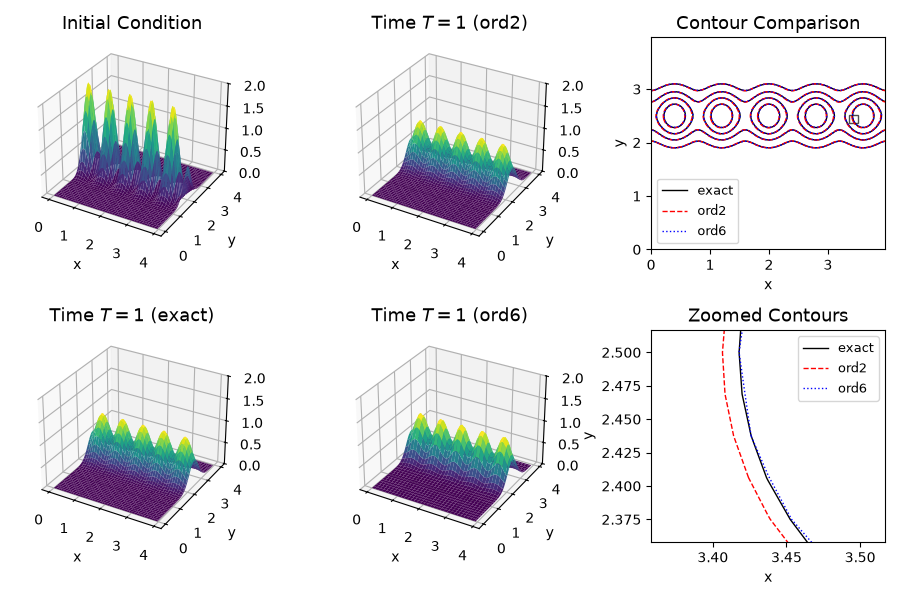}   
\caption{Comparison of QSVT-based methods of order $2$ and $6$ (bottom panels) and exact solution (top right panel) for the 2D advection-diffusion equation with $c = (1,0.5)$, $\nu = 0.02$ and initial data a mixed wave (top left panel, \cref{eq:mixed_wave}).}  
\label{fig:figure6}
\end{figure}

\paragraph{A mixed wave}
We consider mixed advection-diffusion with $c = (0.5,1)$,  $\nu = 0.02$ and the initial function 
\begin{equation}   \label{eq:mixed_wave} 
u_0(x,y) = e^{-7(x-2)^2}(1+\sin(5\pi y/2)),   
\end{equation}
for $(x,y)\in[0,4]^2$ evolved until time $T = 1$ (\cref{fig:figure6} and \cref{tab:table6}).

We compare the methods of order $2$ and $6$ with $14$ and $12$ spatial qubits, respectively. The order $6$ method achieves an accuracy of $\sim 10^{-4}$, which is two orders of magnitude better than the order $2$ method. This comes at a cost of about a factor of $3/2$ as many gates. The second order method is the optimal choice for lower precision simulations. If we added one qubit per spatial dimension for the order $2$ method, we would get an accuracy comparable with the order $6$ method. However, as we observed in the 1d simulations, this would roughly multiply the gate count by a factor of $4$, and the sixth order method becomes the optimal choice.

\begin{table}[htbp]
\centering
\footnotesize
\caption{Data table for the mixed wave simulation in \cref{fig:figure6}.}
\begin{tabularx}{0.75\columnwidth}{cccXXcc}
\toprule
Order & Spatial  & Total  & Error & Success  & 1-qubit  & CNOT  \\
 &  qubits &  qubits &  &  rate &  gates &  gates \\ \midrule
2 & 14 & 19  & 2.146e-02  & 0.0349 & 8372 & 5540 \\ \midrule
6 & 12 & 18  & 2.419e-04  & 0.0348 & 12826 & 9005 \\ \midrule
\end{tabularx}
\label{tab:table6}
\end{table}

\section{Conclusions and future work}
This paper introduced a quantum algorithm based on the quantum singular value transform for the simulation of the linear advection-diffusion equation. In order to apply the QSVT algorithm, we laid out a general framework for constructing block-encodings of finite-difference operators and worked out the construction in detail for symmetric operators of arbitrary order in one dimension. 

The main contributions of this work are the construction and complexity analysis for the concrete problem of advection-diffusion simulation, and an accompanying end-to-end implementation of a number of specific instances of the proposed algorithm. Moreover, our numerical simulations of the algorithm showed that it can be used with $\sim 11-15$ logical qubits and a few thousands of CNOT and $1$-qubit gates depending on the tunable parameters. The theoretical complexity statement and the numerical simulations showed a high order of compatibility and demonstrated the superior performance of the higher order methods, with some minor caveats. 

The results of this paper lay the groundwork for a number of extensions towards quantum simulation of more complex fluid flow models. First, the extension to higher dimensional linear advection-diffusion can be done more efficiently than the basic method used in this paper. More general linear equations can also be considered. These topics have been investigated in the quantum computing literature, but there is typically still a rather large gap between the high-level algorithms constructed and efficient implementation in practice. Of greater interest is the extension to non-linear models such as Korteweg-De Vries equation, the shallow water equations and eventually the compressible Navier-Stokes equations. Here, the shallow water equations is a natural intermediate test case. Finally, the comparison with classical algorithms will be extended to ascertain the efficiency potential of quantum algorithms in full-fledged models of fluid dynamics.

\newpage

\begin{appendix}
\crefalias{section}{appendix}

\section{Quantum circuits for QSVT}     \label{appendix:QSVT}
We give a precise description of the form in which we apply the QSVT algorithm. 

\begin{theorem} \label{thm:QSVT}
Let $(\iota\colon H_k\ra H_n,U\colon H_n\ra H_n)$ be a unitary block-encoding of a Hermitian operator $A\colon H_k\ra H_k$ and let $q\in \R[x]$ be an even or odd polynomial of degree $d$ satisfying $|q(x)|\leq 1$ for all $x\in [-1,1]$. Then there exists an angle sequence (depending only on $q$) $\Phi = (\phi_1,\phi_2,\cdots,\phi_d)$ such that
\[ (\om{QSVT}(U,\Pi,\Phi) \colon (U_\Phi\colon H_1\otimes H_n\ra H_1\otimes H_n, \ket{0}\otimes \iota\colon H_k\ra H_1\otimes H_n )\]
defines a block encoding of $q(A)\colon H_k\ra H_k$, where $\om{QSVT}(U,\Pi,\Phi)$ is defined by the quantum circuit in \cref{fig:QSVT_1}. The resources required for the circuit are given by
\begin{enumerate}[(1)]
\item $d$ applications of $U$ or $U^\dagger$,
\item $2d$ applications of $C_\Pi\om{NOT} \coloneqq X\otimes \Pi + I\otimes (I-\Pi)$ and
\item $(d+2)$ $1$-qubit gates.
\end{enumerate}    
\end{theorem}
\begin{figure}[ht] 
\centering
\begin{quantikz}[column sep = 0.35cm]
&  \gate{H} &  \targ{}  \wire[d][1]{q} & \gate{e^{-i\phi_d \sigma_z}}   &  \targ{} \wire[d][1]{q}   &     &   \targ{}  \wire[d][1]{q} & \gate{e^{-i\phi_{d-1} \sigma_z}}   &  \targ{} \wire[d][1]{q}   &    & \rstick{$\cdots$}   \\
 \setwiretype{b}& \gate{U} & \gate{\Pi} & & \gate{\Pi}  & \gate{U^\dagger}  &  \gate{\Pi} & & \gate{\Pi}  &   & \rstick{$\cdots$}  \\ [0.3cm]
\lstick{$\cdots$} & & \targ{}  \wire[d][1]{q} & \gate{e^{-i\phi_2 \sigma_z}}   &  \targ{} \wire[d][1]{q}   &     &   \targ{}  \wire[d][1]{q} & \gate{e^{-i\phi_{1} \sigma_z}}   &  \targ{} \wire[d][1]{q}   &  \gate{H}  &  \\
\lstick{$\cdots$} \setwiretype{b} & \gate{U} & \gate{\Pi} & & \gate{\Pi}  & \gate{U^\dagger}  &  \gate{\Pi} & & \gate{\Pi}  &   & 
\end{quantikz}
\caption{The quantum circuit $\om{QSVT}(U,\Pi,\Phi)$ for an angle sequence $\Phi = (\phi_1,\cdots,\phi_d)$ with $d$ even.} \label{fig:QSVT_1}
\end{figure} 

\begin{remark} We make some clarifications. 
For a detailed account of the above theorem and its proof, the reader should consult the original reference \cite{QSVT19}. 
\begin{itemize} 
\item The theorem remains true if $A$ is not Hermitian, but then we must apply $q$ to $A$
in the singular value sense defined in \cite[Def.~16]{QSVT19}.
\item The theory explaining the relation between angle sequences and polynomials is called quantum signal processing (QSP). In the above form, the angle sequence associated with a given polynomial is not unique.
\item An efficient and numerically stable way to compute the angle sequence associated with a given polynomial is given in \cite{DongMeng21} with a corresponding MATLAB package QSPACK. Their method is also implemented in the python package pyqsp introduced in \cite{Martyn_etal21}.  
\end{itemize} 
\end{remark}

Polynomials without definite parity can be handled by introducing an additional ancilla qubit and applying \cref{thm:QSVT} for two angle sequences in parallel \cite[Thm.~56]{QSVT19}.  A precise statement is contained in the following corollary.

\begin{corollary}\label{cor:QSVT_parallel} 
Let $(\iota\colon H_k\ra H_n,U\colon H_n\ra H_n)$ be a unitary block-encoding of a Hermitian operator $A\colon H_k\ra H_k$ and let $q = q_1+q_2\in \R[x]$ be a polynomial of degree $d+1$ satisfying $|q_1(x)|, |q_2(x)|\leq 1$ for all $x\in [-1,1]$, where $q_1(x)$ and $q_2(x)$ have opposite parity and we take $\om{deg}q_1<\om{deg}q_2$. Let
\[ \Phi^{(1)} = (\phi_1^{(1)},\cdots,\phi_d^{(1)}) \quad \text{ and } \quad \Phi^{(2)} = (\phi_1^{(2)},\cdots,\phi^{(2)}_d,\phi_{d+1}^{(2)})  \]
be angle sequences associated with $q_1$ and $q_2$, respectively, as given in \cref{thm:QSVT}. Then
\[ (\om{QSVT}(U,\Pi,\Phi_1,\Phi_2)\colon H_1^{\otimes 2}\otimes  H_n\ra H_1^{\otimes 2}\otimes H_n,\; \ket{0}^{\otimes 2}\otimes \iota\colon H_k\ra H_1^{\otimes 2}\otimes H_n )\]
defines a block encoding of $(1/2)q(A)\colon H_k\ra H_k$, where $\om{QSVT}(U,\Pi,\Phi_1,\Phi_2)$ is defined by the quantum circuit in \cref{fig:QSVT_2}. The resources required for the circuit are
\begin{enumerate}[(1)]
\item $d$ applications of $U$ or $U^\dagger$,
\item one application of controlled $U$ (or $U^\dagger$),
\item $2(d+1)$ applications of $C_\Pi\om{NOT}$,
\item $(2d+1)$ $2$-qubit gates and $4$ $1$-qubit gates. 
\end{enumerate} 
\end{corollary}

\begin{figure}[ht] 
\centering
\begin{quantikz}[column sep = 0.35cm, row sep = 0.2cm]
& \gate{H} & \ctrl{2} & & \ctrl{1} &  &  \\
& \gate{H} & & \targ{} \wire[d][1]{q} & \gate{e^{-i\phi^{(2)}_{d+1}\sigma_z}} & \targ{} \wire[d][1]{q}& \\
\setwiretype{b} &&  \gate{U^\dagger} & \gate{\Pi} & & \gate{\Pi} & \\ [0.5cm]
& \ghost{H} & & \octrl{1} & \ctrl{1} & & \rstick{$\cdots$}  \\
& \ghost{H} & \targ{} \wire[d][1]{q} & \gate{e^{-i\phi_{d}^{(1)}\sigma_z}} & \gate{e^{-i\phi_{d}^{(2)}\sigma_z}} & \targ{} \wire[d][1]{q}  &\rstick{$\cdots$} \\
\setwiretype{b} & \gate{U} & \gate{\Pi} & & & \gate{\Pi} & \rstick{$\cdots$} \\ [0.5cm]
\lstick{$\cdots$} & & & \octrl{1} & \ctrl{1} & & \gate{H} &  \\
\lstick{$\cdots$} & & \targ{} \wire[d][1]{q} & \gate{e^{-i\phi_{1}^{(1)}\sigma_z}} & \gate{e^{-i\phi_{1}^{(2)}\sigma_z}} & \targ{} \wire[d][1]{q}  & \gate{H} \\
\lstick{$\cdots$} \setwiretype{b} & \gate{U} & \gate{\Pi} & & & \gate{\Pi} & & 
\end{quantikz}
\caption{The quantum circuit $\om{QSVT}(U,\Pi,\Phi_1,\Phi_2)$ for a pair of angle sequences $\Phi^{(1)} = (\phi_1^{(1)},\ldots,\phi^{(1)}_d)$ and $(\Phi^{(2)} = (\phi^{(2)}_1,\ldots,\phi^{(2)}_{d+1})$ with $d$ odd.}\label{fig:QSVT_2}
\end{figure} 

\begin{remark} To construct the linear combination $q_1(x)+iq_2(x)$ instead of $q_1(x)+q_2(x)$ as in \cref{cor:QSVT_parallel}, it suffices to add an initial gate $S = \ket{0}\bra{0}+i\ket{1}\bra{1}$ to the first qubit in \cref{fig:QSVT_2}.  
\end{remark}

The following lemma contains a useful observation used in the main text. 

\begin{lemma} \label{lem:conjugation_QSVT}
If $F$ is a unitary commuting with $\Pi$, i.e., $F\Pi= \Pi = F$, then
\[ \om{QSVT}(FUF^{-1},\Pi,\Phi) = (I\otimes F)\om{QSVT}(U\dagger,\Pi,\Phi)(I\otimes F^\dagger) .\]
\end{lemma}
\begin{proof} If $F$ commutes with $\Pi$, then $I\otimes F$ commutes with $C_\Pi \om{NOT}$. Therefore, 
if the block-encoding has the form $FUF^{-1}$ in the circuit of \cref{fig:QSVT_1}, every internal occurence of $F$ can be moved two the left over the $C_\Pi\om{NOT}$-gates to cancel the adjacent $F^{-1}$-gate. The result of the cancellations is precisely $(I\otimes F)\om{QSVT}(U\dagger,\Pi,\Phi)(I\otimes F^\dagger)$ as claimed.
\end{proof} 

An analogous result is true for the QSVT version of \cref{cor:QSVT_parallel}.

\section{Block-encoding of finite-difference operators: proofs and an additional example} \label{appendix:block-encoding}
\begin{example} The phase adder $P_{2,3,3} = R_3\circ P_{2,3}$ admits the following quantum circuit representation.
\[ \begin{quantikz}[row sep = 0.15cm, column sep = 0.3cm]
& \ctrl{2} & \ctrl{3} & \ctrl{4} &\ghost{P(\pi/2)} & & \slice{} &&  \\
&  &  &  & \ctrl{1} & \ctrl{2} & \ghost{P(\pi/2)} &&  \\
& \gate{P(\pi/4)} &  & & \gate{P(\pi/2)} & & & \gate{P(3\pi/4)} &   \\
& & \gate{P(\pi/2)} & & & \gate{P(\pi)} & & \gate{P(3\pi/2)} &\\
& & & \gate{P(\pi)} & && & \gate{P(3\pi) } & 
\end{quantikz} \] 
\end{example}

\begin{proof}[Proof of \cref{lem:block_enc_QSVT_complexity}]
It is clear that the circuit uses $2+m+n$ qubits. We estimate the gate complexity for each item in the resource overview given in \cref{cor:QSVT_parallel} as follows. 
\begin{enumerate}[(i)]
\item $D$ applications of $U$ or $U^\dagger$ uses $O(Dmn)$ $2Q$-gates and $O(Dn)$ $1Q$-gates.
\item An application of controlled $U$ (or $U^\dagger$) can naively be implemented by putting controls on all the $2Q$- and $1Q$-gates involved. A doubly controlled $1Q$-gate, say $C^2V$, can be realized with two $\om{CNOT}$ gates, two $C\sqrt{V}$ gates and one $C\sqrt{V}^\dagger$ gate by a standard construction \cite{Barenco_etal95}. Since all of the $2Q$-gates in our construction can be taken to be of this form, $CU$ can be realized with $O(mn)$ gates.
\item $2(D+1)$ applications of $C_\Pi\om{NOT}$. In our case $\Pi = \ket{0}\bra{0}\otimes I_{H_n}$. Therefore, $C_\Pi\om{NOT} = C^m\om{NOT}$ with control state $0 = [0\cdots 0]_2$,
which can be implemented with $O(m^2)$ gates without ancillas \cite{Barenco_etal95} (with ancillas the complexity can be reduced to $O(m)$). The total cost of this item is therefore $O(m^2 D)$.
\item $(2D+1)$ $2Q$-gates and $4$ $1Q$-gates has complexity $O(D)$.
\item Finally, we also have to include a conjugation by the quantum Fourier transform on the $n$-qubit registry. This has complexity $O(n^2)$ (see \cref{lem:conjugation_QSVT}).
\end{enumerate} 
Since we require $m<n$, the resources are dominated by the first item, leading to a total complexity of $O(Dmn)$.
\end{proof}

\section{Complexity analysis: proofs and supporting lemmas} \label{appendix:complexity}
\begin{proof}[Proof of \ref{lem:error_propagation}]
First, by introducing the change of variable $v(t,x) = u(t,x-ct)$, the equation $\dd_tu+c\dd_xu = \nu \dd_x^2u$ is transformed into a pure heat equation $\dd_tv = \nu \dd_x^2v$. With $\rho = d^{-1} \int_0^d u_0(x)dx$, we have the energy estimate (see for instance \cite[13.1]{Quarteroni_etal07})
\begin{align}
||& u_t ||_{L^2} =  ||v_t||_{L^2} \leq ||v_t-\rho||_{L^2} + |\rho| 
 \leq e^{-t\nu \omega^2}||v_0-\rho||_{L^2} + |\rho| \\
 & \leq e^{-t\nu \omega^2}||v_0||_{L^2}+ (1+e^{-t\nu\omega^2})|\rho| =  
 e^{-t\nu \omega^2}||u_0||_{L^2}+ (1+e^{-t\nu\omega^2})|\rho|, \nonumber
\end{align}
proving the first estimate. For the final claim, note that $|c|\leq d^{-1}||u_0||_{L^1} \leq d^{-1/2}||u_0||_{L^2}$, which when combined with the first estimate yields $||u_t||_{L^2} \leq (1+2d^{-1/2})||u_0||$ as required. 
\end{proof}

We need a preparatory lemma.

\begin{lemma} \label{lem:lambda_derivative}
Let $p$ be a positive integer and let $\alpha_j = 2(-1)^{j+1}\frac{(p!)^2}{(p+j)!(p-j)!}$ for $1\leq j\leq p$. Then for $h\in \R$ the following identity holds true
\[ \sum_{j=1}^p \alpha_j \cos(jh) = 1 - 2^{2p}\binom{2p}{p}^{-1} \sin(h/2)^{2p}  .\]
\end{lemma} 
\begin{proof} We compute using the binomial formula
\begin{align*}
(2i\sin & (h/2))^{2p} =(e^{ih/2}-e^{-ih/2})^{2p} =\sum_{j=0}^{2p} \binom{2p}{j} (-1)^j e^{i(j-p)h}\\ 
 &=(-1)^p\left[\binom{2p}{p} + \sum_{j=1}^p \binom{2p}{p-j}(-1)^{j}e^{-ijh} + \sum_{j=1}^p \binom{2p}{p+j}(-1)^{j}e^{ijh}\right] \\
 & = (-1)^p\left[  \binom{2p}{p} + \sum_{j=1}^p2(-1)^j\binom{2p}{p+j}\cos(jh) \right] .
\end{align*}
In the passage to the second line, the sum is first split into three parts $0\leq j\leq p-1$, $j=p$ and $p+1\leq j\leq 2p$, followed by the change of variables $j\mapsto p-j$ in the first sum and $j\mapsto j+p$ in the third sum. The desired formula is obtained by multiplying by $(-1)^p\binom{2p}{p}^{-1}$, rearranging and using the fact that $\alpha_j = (-1)^{j+1}\binom{2p}{p+j}\binom{2p}{p}^{-1}$.
\end{proof} 

\begin{proof}[Proof of \cref{thm:lambda_mu_estimates}] Part $(a)$ is evident from the formulas in \cref{eq:def_lambda_mu}. For part $(b)$, observe first that by \cref{lem:lambda_derivative}
\begin{equation} \label{pf:lambda_der}
\lambda'(k) = \omega \sum_{j=1}^p \alpha_j \cos(\omega k j\Delta x) = \omega\left(1-C\sin(\omega k\Delta x/2)^{2p}\right),
\end{equation} 
where $C\coloneqq 2^{2p}\binom{2p}{p}^{-1}$. Set $f(k) = \omega k-\lambda(k)$. As $f(0) = 0$ and 
\[ f'(k) = \omega-\lambda'(k) = \omega C\sin(\omega k\Delta x/2)^{2p} \geq 0 \quad \text{ for all } k\in \R ,\]
it follows that $f(k)\geq 0 \Leftrightarrow \lambda(k)\leq \omega k$ for $k\geq 0$, proving the first part of $(b)$. Next, notice that
\[ \mu'(k) = 2\omega \sum_{j=1}^p \alpha_j \frac{\sin(\omega k j\Delta x)}{j\Delta x} = 2\omega \lambda(k)  .\]
Hence, for $g(k) \coloneqq \omega^2k^2- \mu(k)$ we obtain $g'(k) = 2\omega(\omega k-\lambda_k)\geq 0$
for $k\geq 0$. As $g(0) = 0$, it follows that $g(k)\geq 0$ or equivalently $\mu(k)\leq \omega^2k^2 $ for all $k\geq 0$. This proves the second part of $(b)$. 

For part $(c)$ it suffices to show that $\mu'(k) = 2\omega \lambda(k) \geq 0$ for $k\in [0,N/2]$.
Using \cref{pf:lambda_der} we find
\[ \lambda''(k) = -(pC\omega^2 \Delta x) \cos(\omega k\Delta x/2)\sin(\omega k\Delta x/2)^{2p-1}.\]
Now, $k\in [0,N/2]$ if and only if $\omega k \Delta x/2 = \pi k/N \in [0,\pi/2]$, from which we can conclude that $\lambda''(k)\leq 0$ for $k\in [0,N/2]$, that is, $\lambda$ is concave. By inspection,
$\lambda(0) = \lambda(N/2) =0$, so by Jensen's inequality it follows that $\lambda(k)\geq 0$ for $k\in [0,N/2]$. 

For part $(d)$, we assume without loss of generality that $k\geq 0$ and estimate
\begin{align*}
\omega k-\lambda_k &= f(k)-f(0) = \int_0^k f'(s)ds  = \omega C\int_0^k \sin(\omega s\Delta x/2)^{2p} ds  \\
&\leq \omega C \int_0^k (\omega s\Delta x/2)^{2p}ds = \frac{(p!)^2}{(2p+1)!}(\omega k)^{2p+1}(\Delta x)^{2p},
\end{align*}
where we have used that $C = 2^{2p}\binom{2p}{p}^{-1}$. Next, by part $(b)$
\[ |\omega^2k^2-\lambda_k^2| =|\omega k + \lambda_k||\omega k-\lambda_k| \leq 2\omega |k||\omega k-\lambda_k|. \]
The second formula follows by combining this with the first estimate. The final estimate can be deduced from the first one as follows
\begin{align*}
\omega^2k^2 -\mu_k &= g(k)-g(0) = \int_0^k g'(s)ds = 2\omega \int_0^k \omega s-\lambda(s)ds \\
&\leq 2\omega \frac{(p!)^2}{(2p+1)!}\omega^{2p+1}(\Delta x)^{2p} \int_0^k s^{2p+1}ds
= \frac{2(p!)^2}{(p+2)!} (\omega k)^{2p+2} (\Delta x)^2
\end{align*}
\end{proof}

\begin{proof}[Proof of \cref{lemma:norm_comparison}]
The basic point is that the coefficients $a_k$ specifying a given $q\in \MC{T}_N$ can be recovered by applying the (unitary) inverse Fourier transform (see \cref{eq:QFT_def})
to $\om{ev}(q)$ and scaling by $N^{-1/2}$. The norm comparison then follows from
\[ d^{-1} ||q||^2_{L^2} = \sum_k |a_k|^2 = N^{-1} ||\om{ev}(q)||^2  .\]
\end{proof}

We need another lemma. 

\begin{lemma}  \label{lem:exp_inequality}
For $z,w\in \C$ it holds true that
\[ |e^z-e^w|\leq  e^{\om{max}(\om{Re}z,\om{Re}w)}|z-w|  .\] 
\end{lemma} 
\begin{proof} Parametrize the line segment $L$ from $w$ to $z$ by $\gamma(t) = w+t(z-w)$, $t\in [0,1]$. Set $b = \om{Re}z$ and $a=\om{Re}w$. We compute
\begin{align*} |e^z-e^w| &= \left|\int_L e^u du\right| = \left|\int_0^1 e^{w+t(z-w)}(z-w)dt\right| \\ 
&\leq \int^1_0 e^{a+t(b-a)}dt |z-w| = \frac{e^b-e^a}{b-a}|z-w| \leq e^{\max(a,b)}|z-w|,
\end{align*}
where the final inequality follows from the mean value theorem. 
\end{proof} 

\begin{proof}[Proof of \cref{thm:L2_estimate}]
Without loss of generality, we work with the normalized $L^2$-norm for which $||e_k(x)||_{L^2} = 1$ for all $k$.
Throughout the proof, all sums are taken from $-N/2$ to $N/2-1$. Set $z_k = -ic\omega kt-\nu \omega^2k^2t$ and $w_k = -ic\lambda_k t-\nu \mu_kt$. By \cref{lem:exp_inequality},
\begin{equation} \label{prf:est1}
||u_t-v_t||_{L^2}^2 = \sum_k |a_k|^2 |e^{z_k}-e^{w_k}|^2 \leq \sum_k |a_k|^2 e^{2\max(\om{Re}z_k,\om{Re}w_k)}|z_k-w_k|^2.    
\end{equation} 
By \cref{thm:lambda_mu_estimates},  $\max(\om{Re}z_k,\om{Re}w_k) = -\nu t \mu_k$ and
\begin{align} \label{pf-DiffEstimate} 
|z_k-w_k|^2 &= (ct)^2|\omega k-\lambda_k|^2+ (\nu t)^2|\omega^2k^2-\mu_k|^2 \\
& \leq (t(\Delta x)^{2p})^2( (cC_p |\omega k|^{2p+1})^2+ (\nu C'_p|\omega k|^{2p+2})^2), \nonumber
\end{align}
for all $|k|\leq N/2$. The desired estimate is obtained by combining this with \cref{prf:est1} as follows
\begin{align*}
 ||u_t-v_t||^2_{L^2} &\leq (t(\Delta x)^{2p})^2\sum_k |a_k|^2e^{-2\nu t\mu_k}\left((cC_p |\omega k|^{2p+1})^2+ (\nu C'_p|\omega k|^{2p+2})^2\right) \\
 &\leq (te^{-\nu t\mu_1}(\Delta x)^{2p})^2\sum_k |a_k|^2\left((cC_p |\omega k|^{2p+1})^2+ (\nu C'_p|\omega k|^{2p+2})^2\right) \\
 &= (te^{-\nu t\mu_1}(\Delta x)^{2p})^2 \left(c^2C_p^2||u_0^{2p+1}||_{L^2}^2 +  \nu^2(C'_p)^2||u_0^{(2p+2)}||_{L^2}^2\right).
\end{align*} 
Here we have used $\mu_1 \leq \mu_k$ for all $0<|k|\leq N/2$ by \cref{thm:lambda_mu_estimates}, and the fact from Fourier analysis that $||u_0^{(m)}||_{2}^2 = \sum_{k\in \Z} |\omega k|^m|a_k|^2$ for $m\geq 0$.
\end{proof}

\section{An additional numerical result}    \label{appendix:numerical_sim} 
\begin{figure}[htbp] 
\centering
\includegraphics[scale=0.56]{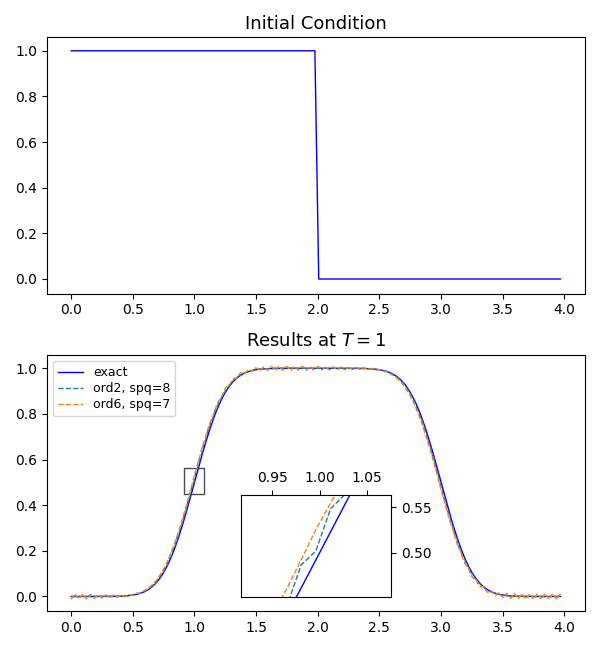}
\caption{Comparison of QSVT-based methods of order $2$ and $6$ and exact solution (bottom panel) for the advection-diffusion equation with $c=1$, $\nu = 0.2$ and initial data a rectangle function (top panel, \cref{eq:rec_func}).} 
\label{fig:figure4}
\end{figure}

Consider a non-smooth rectangle function given by 
\begin{equation} \label{eq:rec_func} 
u_0(x) = \left\{ \begin{array}{ll} 1 & \text{ if } 0\leq x\leq 2  \\ 
                                       0 & \text{ if } 2<x<4 \end{array} \right.
\end{equation} 
for $x\in[0,4]$ subject to advection with $c = 1$ and diffusion with $\nu = 0.02$ for an evolution time of $T = 1$ (\cref{fig:figure4} and \cref{tab:table4}).

We compare the methods of order $2$ and $6$ with $8$ and $7$ spatial qubits, respectively. Both runs achieve an error of order $\sim 10^{-2}$ with the same total number of qubits. The order $2$ method is superior in terms of gates and is therefore favored over the order $6$ method in this case. Since higher order methods are only effective when the initial data is correspondingly regular, these results are not surprising.  

We regard this example as an application of our algorithm to state preparation. The rectangle function is easily prepared with Hadamard gates on the first half of the spatial qubits. By evolving this function with the second order method, we prepare an approximate bump function. The center and slope of the bump function can be controlled through the advection and diffusion parameters, respectively.

\begin{table}[htbp]
\centering
\footnotesize
\caption{Data table for the rectangle function simulations in \cref{fig:figure4}.}
\begin{tabularx}{0.75\columnwidth}{cccXXcc}
\toprule
Order & Spatial  & Total  & Error & Success  & 1-qubit  & CNOT  \\
 &  qubits &  qubits &  &  rate &  gates &  gates \\ \midrule
2 & 8 & 12  & 3.323e-02  & 0.2218 & 8607 & 5970 \\ \midrule
6 & 7 & 12  & 3.376e-02  & 0.2219 & 13187 & 9569 \\ \midrule
\end{tabularx}
\label{tab:table4}
\end{table}

\end{appendix}

\newpage

\bibliographystyle{siamplain}
\bibliography{biblio}

@misc{qiskit24,
  title        = {Quantum computing with {Q}iskit},
  author       = {Javadi-Abhari, Ali and Treinish, Matthew and Krsulich, Kevin and Wood, Christopher J. and Lishman, Jake and Gacon, Julien and Martiel, Simon and Nation, Paul D. and Bishop, Lev S. and Cross, Andrew W. and Johnson, Blake R. and Gambetta, Jay M.},
  year         = {2024},
  doi          = {10.48550/arXiv.2405.08810},
  eprint       = {2405.08810},
  archivePrefix= {arXiv},
  primaryClass = {quant-ph},
}

@misc{gitrep_GOH,
author = {Helle, Gard Olav and Ousager, Anna Bomme},
title = {\url{https://github.com/GOHelle/QC_advection-diffusion_sim}},
year= {2025}
}

@article{Xie_etal24,
author = {Xie, Hao-Nan and Wei, Shijie and Yang, Fan and Wang, Zheng-An and Chen, Chi-Tong and Fan, Heng and Long, Gui},
year = {2024},
month = {05},
pages = {},
title = {Probabilistic imaginary-time evolution algorithm based on nonunitary quantum circuits},
volume = {109},
journal = {Physical Review A},
doi = {10.1103/PhysRevA.109.052414}
}

@article{Hu_etal24,
author = {Hu, Junpeng and Jin, Shi and Liu, Nana and Zhang, Lei},
year = {2024},
month = {12},
pages = {1563},
title = {Quantum Circuits for partial differential equations via Schrödingerisation},
volume = {8},
journal = {Quantum},
doi = {10.22331/q-2024-12-12-1563}
}

@article{Ingelmann_etal24,
author = {Ingelmann, Julia and Bharadwaj, Sachin and Pfeffer, Philipp and Sreenivasan, Katepalli and Schumacher, Jörg},
year = {2024},
month = {07},
pages = {106369},
title = {Two quantum algorithms for solving the one-dimensional advection-diffusion equation},
volume = {281},
journal = {Computers \& Fluids},
doi = {10.1016/j.compfluid.2024.106369}
}

@article{Over_etal25,
author = {Over, Paul and Bengoechea, Sergio and Brearley, Peter and Laizet, Sylvain and Rung, Thomas},
year = {2025},
month = {06},
pages = {},
title = {Quantum algorithm for the advection-diffusion equation by direct block encoding of the time-marching operator},
volume = {112},
journal = {Physical Review A},
doi = {10.1103/d8hb-fv93}
}

@article{BrearleySylvain24,
author = {Brearley, Peter and Laizet, Sylvain},
year = {2024},
month = {07},
pages = {},
title = {Quantum algorithm for solving the advection equation using Hamiltonian simulation},
volume = {110},
journal = {Physical Review A},
doi = {10.1103/PhysRevA.110.012430}
}

@article{Lubasch_etal25,
  title = {Quantum circuits for partial differential equations in Fourier space},
  author = {Lubasch, Michael and Kikuchi, Yuta and Wright, Lewis and Mc Keever, Conor},
  journal = {Phys. Rev. Res.},
  volume = {7},
  issue = {4},
  pages = {043326},
  numpages = {20},
  year = {2025},
  month = {Dec},
  publisher = {American Physical Society},
  doi = {10.1103/tbzc-w9x8},
  url = {https://link.aps.org/doi/10.1103/tbzc-w9x8}
}

@ARTICLE{Novikau24,
       author = {{Novikau}, Ivan and {Joseph}, Ilon},
        title = "{Quantum algorithm for the advection-diffusion equation and the Koopman-von Neumann approach to nonlinear dynamical systems}",
      journal = {arXiv e-prints},
         year = 2024,
        month = oct,
          eid = {arXiv:2410.03985},
        pages = {arXiv:2410.03985},
          doi = {10.48550/arXiv.2410.03985},
archivePrefix = {arXiv},
       eprint = {2410.03985},
 primaryClass = {physics.comp-ph},
       adsurl = {https://ui.adsabs.harvard.edu/abs/2024arXiv241003985N}
}

@inproceedings{QSVT19,
author = {Gily\'{e}n, Andr\'{a}s and Su, Yuan and Low, Guang Hao and Wiebe, Nathan},
title = {Quantum singular value transformation and beyond: exponential improvements for quantum matrix arithmetics},
year = {2019},
isbn = {9781450367059},
publisher = {Association for Computing Machinery},
address = {New York, NY, USA},
url = {https://doi.org/10.1145/3313276.3316366},
doi = {10.1145/3313276.3316366},
booktitle = {Proceedings of the 51st Annual ACM SIGACT Symposium on Theory of Computing},
pages = {193–204},
numpages = {12},
location = {Phoenix, AZ, USA},
series = {STOC 2019}
}

@ARTICLE{Martyn_etal21,
author = {{Martyn}, John M. and {Rossi}, Zane M. and {Tan}, Andrew K. and {Chuang}, Isaac L.},
title = "{Grand Unification of Quantum Algorithms}",
journal = {PRX Quantum},
 year = 2021,
month = dec,
volume = {2},
number = {4},
eid = {040203},
pages = {040203},
doi = {10.1103/PRXQuantum.2.040203},
archivePrefix = {arXiv},
eprint = {2105.02859},
primaryClass = {quant-ph}
}

@book{QAlg25,
author = {Dalzell, Alexander and McArdle, Sam and Berta, Mario and Bienias, Przemyslaw and Chen, Chi-Fang and Gilyén, András and Hann, Connor and Kastoryano, Michael and Khabiboulline, Emil and Kubica, Aleksander and Salton, Grant and Wang, Samson and Brandão, Fernando},
year = {2025},
month = {05},
pages = {},
title = {Quantum Algorithms: A Survey of Applications and End-to-end Complexities},
isbn = {9781009639644},
doi = {10.1017/9781009639651},
publisher = {Cambridge University Press}
}

@article{Motlagh_Wiebe24,
author = {Motlagh, Danial and Wiebe, Nathan},
year = {2024},
month = {06},
pages = {},
title = {Generalized Quantum Signal Processing},
volume = {5},
journal = {PRX Quantum},
doi = {10.1103/PRXQuantum.5.020368}
}

@article{LC17,
  title = {Optimal Hamiltonian Simulation by Quantum Signal Processing},
  author = {Low, Guang Hao and Chuang, Isaac L.},
  journal = {Phys. Rev. Lett.},
  volume = {118},
  issue = {1},
  pages = {010501},
  numpages = {5},
  year = {2017},
  month = {Jan},
  publisher = {American Physical Society},
  doi = {10.1103/PhysRevLett.118.010501},
  url = {https://link.aps.org/doi/10.1103/PhysRevLett.118.010501}
}

@book{NielsenChuang10,
    place={Cambridge},
    title={Quantum Computation and Quantum Information: 10th Anniversary Edition},
    DOI={10.1017/CBO9780511976667},
    publisher={Cambridge University Press},
    author={Nielsen, Michael A. and Chuang, Isaac L.},
    year={2010}}

@article{HHL09,
	doi = {10.1103/physrevlett.103.150502},
	url = {https://doi.org/10.1103%2Fphysrevlett.103.150502},
	year = 2009,
	month = {oct},
	publisher = {American Physical Society ({APS})},
	volume = {103},
	number = {15},
	author = {Aram W. Harrow and Avinatan Hassidim and Seth Lloyd},
	title = {Quantum Algorithm for Linear Systems of Equations},
	journal = {Physical Review Letters}
}

@article{Childs_Kothari_Somma15,
  title={Quantum Algorithm for Systems of Linear Equations with Exponentially Improved Dependence on Precision},
  author={Andrew M. Childs and Robin Kothari and Rolando D. Somma},
  journal={SIAM J. Comput.},
  year={2015},
  volume={46},
  pages={1920-1950},
  url={https://api.semanticscholar.org/CorpusID:3834959}
}

@article{DongMeng21,
author = {Dong, Yulong and Meng, Xiang and Whaley, K. and Lin, Lin},
year = {2021},
month = {04},
pages = {},
title = {Efficient phase-factor evaluation in quantum signal processing},
volume = {103},
journal = {Physical Review A},
doi = {10.1103/PhysRevA.103.042419}
}

@misc{Draper00,
      title={Addition on a Quantum Computer}, 
      author={Thomas G. Draper},
      year={2000},
      eprint={quant-ph/0008033},
      archivePrefix={arXiv},
      primaryClass={quant-ph},
      url={https://arxiv.org/abs/quant-ph/0008033}, 
}

@article{Camps_etal24,
    author = "Camps, Daan and Lin, Lin and Van Beeumen, Roel and Yang, Chao",
    title = "{Explicit Quantum Circuits for Block Encodings of Certain Sparse Matrices}",
    eprint = "2203.10236",
    archivePrefix = "arXiv",
    primaryClass = "quant-ph",
    doi = "10.1137/22M1484298",
    journal = "SIAM J. Matrix Anal. Appl.",
    volume = "45",
    number = "1",
    pages = "801--827",
    year = "2024"
}

@misc{Aggarwal_Alman22,
      title={Optimal-Degree Polynomial Approximations for Exponentials and Gaussian Kernel Density Estimation}, 
      author={Amol Aggarwal and Josh Alman},
      year={2022},
      eprint={2205.06249},
      archivePrefix={arXiv},
      primaryClass={cs.CC},
      url={https://arxiv.org/abs/2205.06249}, 
}

@misc{Low_Chuang17,
      title={Hamiltonian Simulation by Uniform Spectral Amplification}, 
      author={Guang Hao Low and Isaac L. Chuang},
      year={2017},
      eprint={1707.05391},
      archivePrefix={arXiv},
      primaryClass={quant-ph},
      url={https://arxiv.org/abs/1707.05391}, 
}

@article{Sun_etal21,
  title={Asymptotically Optimal Circuit Depth for Quantum State Preparation and General Unitary Synthesis},
  author={Xiaoming Sun and Guojing Tian and Shuai Yang and Pei Yuan and Shengyu Zhang},
  journal={IEEE Transactions on Computer-Aided Design of Integrated Circuits and Systems},
  year={2021},
  volume={42},
  pages={3301-3314},
  url={https://api.semanticscholar.org/CorpusID:237048488}
}

@article{Barenco_etal95,
  title={Elementary gates for quantum computation.},
  author={Adriano Barenco and Charles H. Bennett and Richard Cleve and David P. DiVincenzo and Norman Margolus and Peter W. Shor and T. Sleator and John A. Smolin and Harald Weinfurter},
  journal={Physical review. A, Atomic, molecular, and optical physics},
  year={1995},
  volume={52 5},
  pages={
          3457-3467
        },
  url={https://api.semanticscholar.org/CorpusID:8764584}
}

@article{Childs_Wiebe12,
author = {Childs, Andrew and Wiebe, Nathan},
year = {2012},
month = {02},
pages = {},
title = {Hamiltonian Simulation Using Linear Combinations of Unitary Operations},
volume = {12},
journal = {Quantum Information and Computation},
doi = {10.26421/QIC12.11-12-1}
}

@article{LiNiYing23,
author = {Li, Haoya and Ni, Hongkang and Ying, Lexing},
year = {2023},
month = {06},
pages = {1031},
title = {On efficient quantum block encoding of pseudo-differential operators},
volume = {7},
journal = {Quantum},
doi = {10.22331/q-2023-06-02-1031}
}

@article {Epstein05,
    AUTHOR = {Epstein, Charles L.},
     TITLE = {How well does the finite {F}ourier transform approximate the
              {F}ourier transform?},
   JOURNAL = {Comm. Pure Appl. Math.},
  FJOURNAL = {Communications on Pure and Applied Mathematics},
    VOLUME = {58},
      YEAR = {2005},
    NUMBER = {10},
     PAGES = {1421--1435},
      ISSN = {0010-3640,1097-0312},
   MRCLASS = {42A38},
  MRNUMBER = {2162785},
MRREVIEWER = {J.\ M. H. Peters},
       DOI = {10.1002/cpa.20064},
       URL = {https://doi.org/10.1002/cpa.20064},
}

@book {HandbookFunc,
    AUTHOR = {Abramowitz, Milton and Stegun, Irene A.},
     TITLE = {Handbook of mathematical functions with formulas, graphs, and
              mathematical tables},
    SERIES = {National Bureau of Standards Applied Mathematics Series},
    VOLUME = {No. 55},
      NOTE = {For sale by the Superintendent of Documents},
 PUBLISHER = {U. S. Government Printing Office, Washington, DC},
      YEAR = {1964},
     PAGES = {xiv+1046},
   MRCLASS = {33.00 (65.05)},
  MRNUMBER = {167642},
MRREVIEWER = {D.\ H.\ Lehmer},
}

@book {Quarteroni_etal07,
    AUTHOR = {Quarteroni, Alfio and Sacco, Riccardo and Saleri, Fausto},
     TITLE = {Numerical mathematics},
    SERIES = {Texts in Applied Mathematics},
    VOLUME = {37},
   EDITION = {Second},
 PUBLISHER = {Springer-Verlag, Berlin},
      YEAR = {2007},
     PAGES = {xviii+655},
      ISBN = {978-3-540-34658-6; 3-540-34658-9},
   MRCLASS = {65-01},
  MRNUMBER = {2265914},
       DOI = {10.1007/b98885},
       URL = {https://doi.org/10.1007/b98885},
}

@article{Li05_FD,
       author = {{Li}, Jianping},
        title = "{General explicit difference formulas for numerical differentiation}",
      journal = {Journal of Computational and Applied Mathematics},
         year = 2005,
        month = nov,
       volume = {183},
        pages = {29-52},
       adsurl = {https://ui.adsabs.harvard.edu/abs/2005JCoAM.183...29L}
}

@article{ZhouWang17,
author = {Zhou, Sisi and Wang, Jingbo},
year = {2017},
month = {05},
pages = {160906},
title = {Efficient quantum circuits for dense circulant and circulant like operators},
volume = {4},
journal = {Royal Society Open Science},
doi = {10.1098/rsos.160906}
}

@misc{Jennings_etal25a,
      title={Quantum algorithms for general nonlinear dynamics based on the Carleman embedding}, 
      author={David Jennings and Kamil Korzekwa and Matteo Lostaglio and Andrew T Sornborger and Yigit Subasi and Guoming Wang},
      year={2025},
      eprint={2509.07155},
      archivePrefix={arXiv},
      primaryClass={quant-ph},
      url={https://arxiv.org/abs/2509.07155}, 
}

@article{Joseph20,
  title = {Koopman--von Neumann approach to quantum simulation of nonlinear classical dynamics},
  author = {Joseph, Ilon},
  journal = {Phys. Rev. Res.},
  volume = {2},
  issue = {4},
  pages = {043102},
  numpages = {17},
  year = {2020},
  month = {Oct},
  publisher = {American Physical Society},
  doi = {10.1103/PhysRevResearch.2.043102},
  url = {https://link.aps.org/doi/10.1103/PhysRevResearch.2.043102}
}

@article{Xue_etal25,
author = {Cheng Xue and Xiao-Fan Xu and Xi-Ning Zhuang and Tai-Ping Sun
and Yun-Jie Wang and Ming-Yang Tan and Chuang-Chao Ye
and Huan-Yu Liu and Yu-Chun Wu and Zhao-Yun Chen
and Guo-Ping Guo},
title = {Quantum homotopy analysis method with quantum-compatible
linearization for nonlinear partial differential equations},
journal = {Science China Physics, Mechanics \& Astronomy},
year = {2025},
volume = {68},
number = {10},
pages = {104702},
doi = {10.1007/s11433-024-2584-2}
}

@article{Aaronson04_PostBQP,
  title={Quantum computing, postselection, and probabilistic polynomial-time},
  author={Scott Aaronson},
  journal={Proceedings of the Royal Society A: Mathematical, Physical and Engineering Sciences},
  year={2004},
  volume={461},
  pages={3473 - 3482},
  url={https://api.semanticscholar.org/CorpusID:1770389}
}

@article{Krovi23,
  doi = {10.22331/q-2023-02-02-913},
  url = {https://doi.org/10.22331/q-2023-02-02-913},
  title = {Improved quantum algorithms for linear and nonlinear differential equations},
  author = {Krovi, Hari},
  journal = {{Quantum}},
  issn = {2521-327X},
  publisher = {{Verein zur F{\"{o}}rderung des Open Access Publizierens in den Quantenwissenschaften}},
  volume = {7},
  pages = {913},
  month = feb,
  year = {2023}
}

@article{Liu_etal21,
author = {Liu, Jin-Peng and Kolden, Herman and Krovi, Hari and Loureiro, Nuno and Trivisa, Konstantina and Childs, Andrew},
year = {2021},
month = {08},
pages = {e2026805118},
title = {Efficient quantum algorithm for dissipative nonlinear differential equations},
volume = {118},
journal = {Proceedings of the National Academy of Sciences},
doi = {10.1073/pnas.2026805118}
}

@article{Kivlichan_2017,
doi = {10.1088/1751-8121/aa77b8},
url = {https://doi.org/10.1088/1751-8121/aa77b8},
year = {2017},
month = {jun},
publisher = {IOP Publishing},
volume = {50},
number = {30},
pages = {305301},
author = {Kivlichan, Ian D and Wiebe, Nathan and Babbush, Ryan and Aspuru-Guzik, Alán},
title = {Bounding the costs of quantum simulation of many-body physics in real space},
journal = {Journal of Physics A: Mathematical and Theoretical}
}

@article{ChildsOstrander21,
author = {Childs, Andrew and Ostrander, Aaron},
year = {2021},
month = {11},
pages = {574},
title = {High-precision quantum algorithms for partial differential equations},
volume = {5},
journal = {Quantum},
doi = {10.22331/q-2021-11-10-574}
}

@ARTICLE{AnChildsLin23,
       author = {{An}, Dong and {Childs}, Andrew M. and {Lin}, Lin},
        title = "{Quantum algorithm for linear non-unitary dynamics with near-optimal dependence on all parameters}",
      journal = {arXiv e-prints},
         year = 2023,
        month = dec,
          eid = {arXiv:2312.03916},
        pages = {arXiv:2312.03916},
          doi = {10.48550/arXiv.2312.03916},
archivePrefix = {arXiv},
       eprint = {2312.03916},
 primaryClass = {quant-ph},
       adsurl = {https://ui.adsabs.harvard.edu/abs/2023arXiv231203916A}
}

@ARTICLE{Costa_etal25,
       author = {{Costa}, Pedro C.~S. and {Schleich}, Philipp and {Morales}, Mauro E.~S. and {Berry}, Dominic W.},
        title = "{Further improving quantum algorithms for nonlinear differential equations via higher-order methods and rescaling}",
      journal = {npj Quantum Information},
         year = 2025,
        month = aug,
       volume = {11},
       number = {1},
          eid = {141},
        pages = {141},
          doi = {10.1038/s41534-025-01084-z},
       adsurl = {https://ui.adsabs.harvard.edu/abs/2025npjQI..11..141C}
}

@article{BerryCosta22,
  title={Quantum algorithm for time-dependent differential equations using Dyson series},
  author={Dominic W. Berry and Pedro C. S. Costa},
  journal={Quantum},
  year={2022},
  url={https://api.semanticscholar.org/CorpusID:254366545}
}

@article{Berry_etal17,
  title={Quantum Algorithm for Linear Differential Equations with Exponentially Improved Dependence on Precision},
  author={Dominic W. Berry and Andrew M. Childs and Aaron Ostrander and Guoming Wang},
  journal={Communications in Mathematical Physics},
  year={2017},
  volume={356},
  pages={1057 - 1081},
  url={https://api.semanticscholar.org/CorpusID:119422776}
}

@article{Bauer_etal15,
  author       = {Peter Bauer and Alan Thorpe and Gilbert Brunet},
  title        = {The quiet revolution of numerical weather prediction},
  journal      = {Nature},
  volume       = {525},
  number       = {7567},
  pages        = {47--55},
  month        = sep,
  year         = {2015},
  doi          = {10.1038/nature14956},
}

@article{Schulthess_etal18,
  author       = {Thomas C. Schulthess and Peter Bauer and Nils P. Wedi and Oliver Fuhrer and Torsten Hoefler and Christoph Schär},
  title        = {Reflecting on the Goal and Baseline for Exascale Computing: A Roadmap Based on Weather and Climate Simulations},
  journal      = {Computing in Science \& Engineering},
  volume       = {21},
  number       = {1},
  pages        = {30--41},
  year         = {2018},
  doi          = {10.1109/MCSE.2018.2888788},
}
\end{document}